\pdfoutput=1

\documentclass[%
reprint,
superscriptaddress,
showkeys,
 amsmath,amssymb,
 aps,
prd,
]{revtex4-1}

\usepackage{graphicx}
\usepackage{dcolumn}
\usepackage{bm}
\usepackage{hyperref}
\usepackage{multirow}

\begin{document}

\preprint{Physical Review D}

\title{Reconstructing the direction of reactor antineutrinos via electron scattering in Gd-doped water Cherenkov detectors}

\author{D. Hellfeld}
\thanks{Corresponding author} 
\email[Email: ] {dhellfeld@berkeley.edu}
\thanks{Tel.: +1 949 680 9345}
\affiliation{Department of Nuclear Engineering, University of California, Berkeley, Berkeley, CA 94720}
\author{S. Dazeley}
\thanks{Second author}
\email[Email: ]{dazeley2@llnl.gov}
\thanks{Tel.: +1 925 423 4792.}
\affiliation{Lawrence Livermore National Laboratory, Livermore, CA 94550}
\author{A. Bernstein}
\affiliation{Lawrence Livermore National Laboratory, Livermore, CA 94550}
\author{C. Marianno}
\affiliation{Department of Nuclear Engineering, Texas A\&M University, College Station, TX 77843}

\date{\today}

\begin{abstract}
The potential of elastic antineutrino-electron scattering in a Gd-doped water Cherenkov detector to determine the direction of a nuclear reactor antineutrino flux was investigated using the recently proposed WATCHMAN antineutrino experiment as a baseline model. The expected scattering rate was determined assuming a 13-km standoff from a 3.758-GWt light water nuclear reactor and the detector response was modeled using a Geant4-based simulation package. Background was estimated via independent simulations and by scaling published measurements from similar detectors. Background contributions were estimated for solar neutrinos, misidentified reactor-based inverse beta decay interactions, cosmogenic radionuclides, water-borne radon, and gamma rays from the photomultiplier tubes (PMTs), detector walls, and surrounding rock. We show that with the use of low background PMTs and sufficient fiducialization, water-borne radon and cosmogenic radionuclides pose the largest threats to sensitivity. Directional sensitivity was then analyzed as a function of radon contamination, detector depth, and detector size. The results provide a list of experimental conditions that, if satisfied in practice, would enable antineutrino directional reconstruction at  3$\sigma$ significance in large Gd-doped water Cherenkov detectors with greater than 10-km standoff from a nuclear reactor.
\end{abstract}



\maketitle


\section{\label{sec:intro}Introduction}

Near-field ($<$ 100 m) monitoring of nuclear reactors via measurements of the antineutrino flux and energy spectrum  has been demonstrated using cubic meter scale liquid scintillator antineutrino detectors such as \cite{Bowden, Mikaelyan, Bernstein2008}. With such measurements, reactor characteristics such as the operational status (on/off), relative power output, and the evolution of the fissionable isotopics in the fuel (burnup) could be determined. The success of these detectors has spurred research in much larger detectors in order to increase both sensitivity and standoff distance \cite{Bernstein_2, Lasserre}. Such detectors could potentially be used as a tool in the nuclear safeguards regime set forth by the International Atomic Energy Agency (IAEA) to reduce the effort needed to conduct physical inspections inside of declared reactor facilities, to monitor facilities in which inspectors do not have access, or to either exclude or search for the presence of clandestine reactors in suspected locations.

Kiloton and megaton scale Gd-doped water Cherenkov antineutrino detectors (WCDs), such as the recently proposed WATer CHerenkov Monitor of ANtineutrinos project (WATCHMAN) \cite{Bernstein}, are being investigated for medium to long range ($>$ 10 km) remote monitoring of nuclear reactors. These detectors utilize the coincident detection of the positron and neutron from the inverse beta decay (IBD) interaction ($\bar{\nu}_{e}+p\rightarrow n + e^{+}$) to determine both the flux and energies of the incident antineutrinos. Water is an attractive option when scaling to such large detector sizes primarily due to both cost and environmental factors; and gadolinium is added (typically 0.1\% by weight) to significantly increase both the neutron-tagging efficiency ($\sim$85\%) and capture energy release ($\sim$8 MeV). In this work, we analyze whether, in addition to the rate and energy, these detectors can determine the direction of the incident antineutrinos. Directional sensitivity might prove crucial in instances where multiple reactors are located nearby, or if a clandestine reactor has been confirmed via the IBD signal, directionality could be used in conjunction with other measurements, such as satellite imagery, to determine the location of the reactor. Once the location is known, other methods could be employed to further characterize the reactor.  

Event-by-event reconstruction of the antineutrino direction via IBD in hydrogenous media requires knowledge of the neutron momentum vector within a few recoils following its production. This method of directional reconstruction has not yet been accomplished for reactor antineutrinos in any medium. In liquid scintillator detectors, CHOOZ \cite{Apollonio} has shown that a partial and stochastic knowledge of the direction of an incoming antineutrino flux may be gained over time by reconstructing the relative positions of the positron and neutron thermal capture interaction vertices from an ensemble of IBD interactions. WCDs, however, presently do not possess the spatial resolution or sensitivity to do this. In this paper, we investigate whether an alternative interaction, elastic electron scattering (ES), can be used to determine the direction of a reactor antineutrino flux incident upon a WCD. The ES interaction ($\bar{\nu_e} + e^-$$\rightarrow \bar{\nu_e} + e^-$) is highly directional, meaning the electrons are primarily scattered with a small scattering angle relative to the incident antineutrino. Thus, in principle, the direction of the incident antineutrino flux can be determined via directional reconstructions of an ensemble of scattered electrons.

\subsection{\label{sec:nueinteractions}Antineutrino-electron scattering}
Neglecting the neutrino mass, the elastic antineutrino-electron scattering cross-section in the laboratory frame including both the neutral and charged current components can be written as
\hypertarget{eq1}{}
\begin{align} 
\sigma(E_{\bar{\nu}_{e}})&=\Big(\frac{G^{2}_\text{F}m_{e}E_{\bar{\nu}_{e}}}{6\pi}\Big)\Big[(1+2\sin^{2}\theta_W)^2+12\sin^{4}\theta_W\Big] \nonumber \\
&\simeq (7.8 \times 10^{-45})m_{e}E_{\bar{\nu}_{e}} \hspace{5pt} \text{cm}^{2}\text{ MeV\textsuperscript{-2}}\,,
\end{align}

\noindent where $G_\text{F}$ is the Fermi coupling constant [= $ 1.166364 \times 10$\textsuperscript{-5}  GeV\textsuperscript{-2} ($\hbar c$)\textsuperscript{3}] and $\theta_W$ is the Weinberg mixing angle ($\sin^{2}\theta_W$ $\simeq 0.23$) \cite{Giunti_Kim}. Though the ES cross-section is much smaller than IBD, note that the nuclear reactor antineutrino flux is concentrated at low energies, where the interaction cross-section difference is smallest. Water also presents five times as many ES targets as IBD per water molecule (10 $e^-$ vs. 2 quasi-free protons) [see Fig.\,\hyperlink{fig1}{1(a)}].

From energy and momentum conservation in the laboratory frame, it can be shown that the kinetic energy of the scattered electron $T_e$, is given by
\hypertarget{eq2}{}
\begin{equation} \tag{2}
{T_{e}}(\theta,E_{\bar{\nu}_{e}}\big)=\frac{2m_{e}E^{2}_{\bar{\nu}_{e}}\cos^{2}\theta}{{\big(m_{e}+E_{\bar{\nu}_{e}}}\big)^2-E^{2}_{\bar{\nu}_{e}}\cos^{2}\theta}\,,
\end{equation}

\noindent where $\theta$ is the angle between the incident antineutrino and the scattered electron \cite{Giunti_Kim}. Using this, the differential cross-section as a function of the cosine of the scattering angle can be expressed by 
\hypertarget{eq3}{}
\begin{multline} \tag{3}
\frac{\text{d}\sigma}{\text{d}\cos\theta}\big(\theta,E_{\bar{\nu}_{e}}\big)=\frac{4\sigma_0E^{2}_{\bar{\nu}_{e}}M^2\cos\theta}{(M^2-E^{2}_{\bar{\nu}_{e}}\cos^{2}\theta)^2}\cdot\Bigg[g^{2}_{1}+g^{2}_{2}\cdot
\\
\Bigg(1 -\frac{2m_{e}E_{\bar{\nu}_{e}}\cos^{2}\theta}{M^2-E^{2}_{\bar{\nu}_{e}}\cos^{2}\theta}\Bigg)^{2}-\frac{2m^{2}_{e}g_{1}g_{2}\cos^{2}\theta}{M^2-E^{2}_{\bar{\nu}_{e}}\cos^{2}\theta}\Bigg]\,,
\end{multline}

\noindent where $\sigma_0$ = 88.06 $\times$ 10\textsuperscript{-46} cm\textsuperscript{2}, $M = m_{e}+E_{\bar{\nu}_{e}}$, $g_{1}=\frac{1}{2}\big(g_{V}-g_{A}\big)$, and $g_{2}=\frac{1}{2}\big(g_{V}+g_{A}\big)$ where $g_V$ and $g_A$ are the weak vector and weak axial-vector coupling constants, respectively \cite{Giunti_Kim}. The differential cross-section is plotted in Fig.\,\hyperlink{fig1}{1(b)} for several incident antineutrino energies. The trend of the cross-section to increase towards $\cos\theta = 1$ reveals that the scattered electrons are primarily scattered in the direction of the incident antineutrinos. Note that the effect becomes more apparent as the incident antineutrino energy increases.

\begin{figure}[!htb]
\hypertarget{fig1}{}
\centering
\begin{tabular}{c}
\includegraphics[width=245pt]{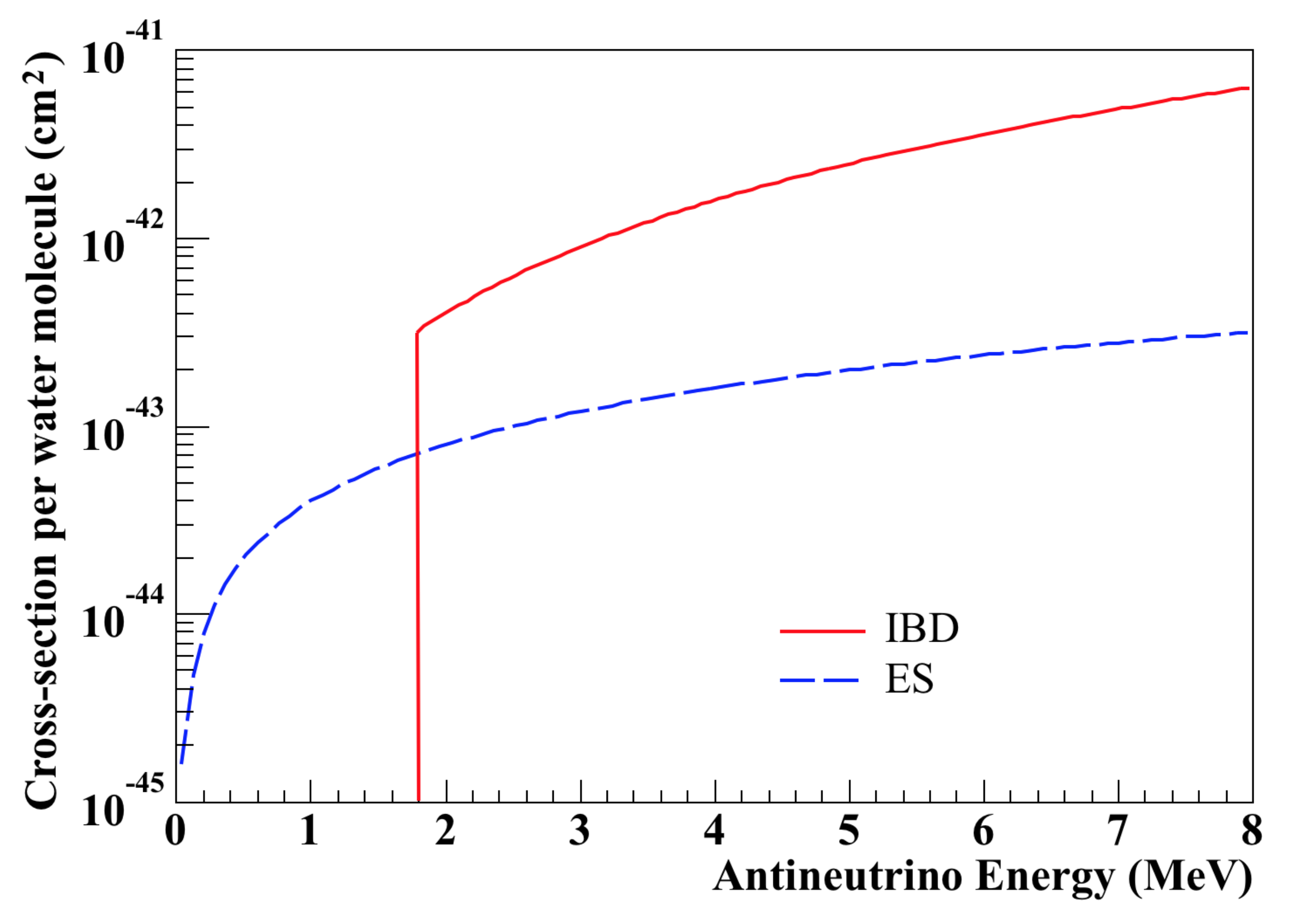} \\
\footnotesize (a) \\
\includegraphics[width=245pt]{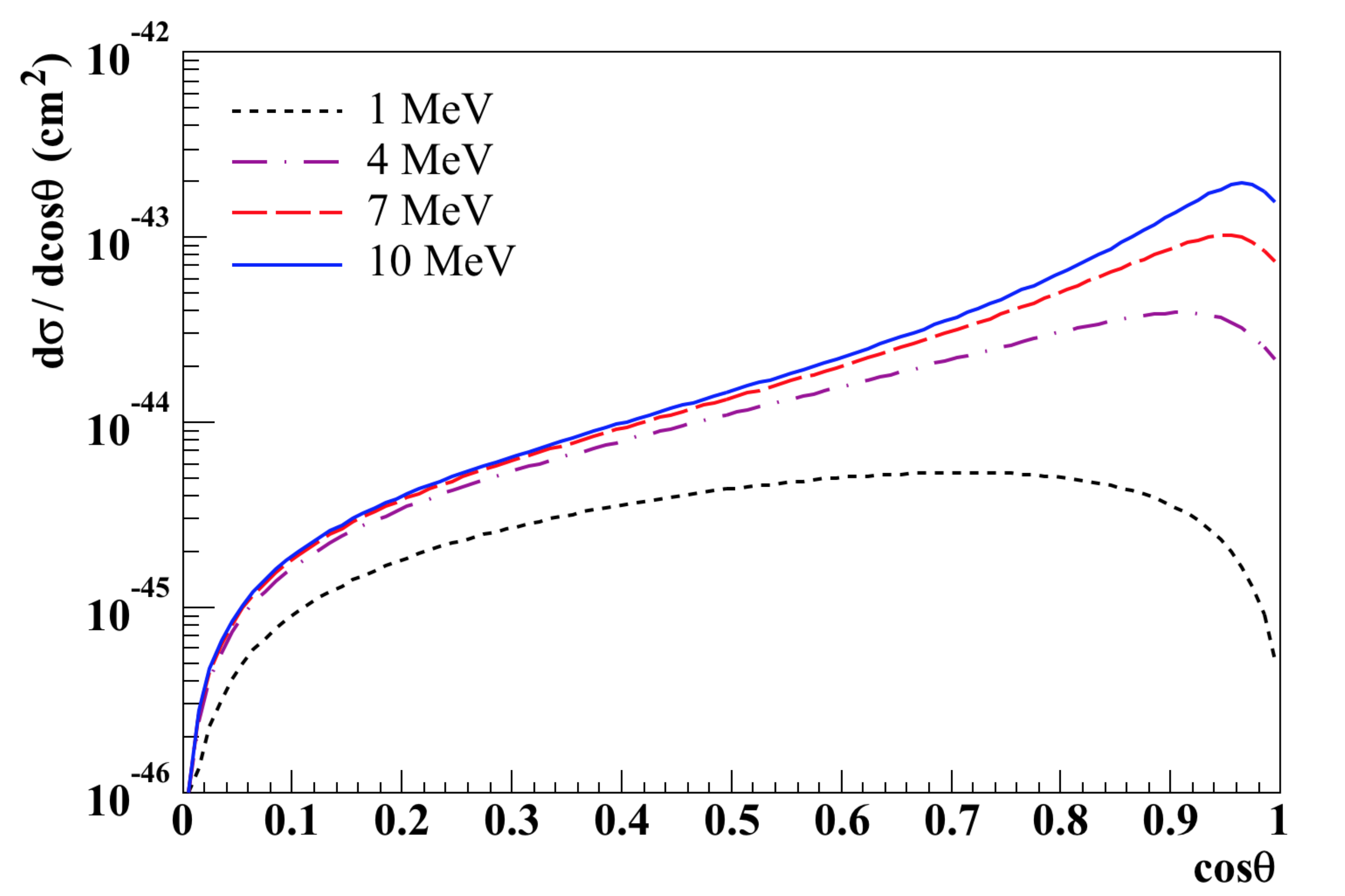} \\
\footnotesize (b)
\end{tabular}
\caption{(a) ES and IBD cross-sections per water molecule as functions of incident antineutrino energy. Note the 1.8-MeV energy threshold for IBD. (b) Antineutrino-electron scattering differential cross-section as a function of the cosine of the scattering angle $\theta$.}
\end{figure}

\subsection{\label{sec:energyspectrum}Reactor antineutrino energy spectrum}

The fission of uranium and plutonium inside of nuclear reactor systems produce neutron-rich fission fragment pairs, which beta decay six times on average before reaching stability. Each one of these decays will produce an antineutrino with a continuum of possible energies. Therefore, experiments and simulations are used to study both the production and subsequent decay of fission products in critical nuclear reactor systems to understand the reactor antineutrino energy spectrum. As shown by \cite{Vogel_Engel}, the number of antineutrinos produced per fission per MeV can be modeled for a particular fissionable isotope by 
\hypertarget{eq4}{}
\begin{equation} \tag{4}
\phi(E_{\bar{\nu}_{e}}\big)=\text{exp}\bigg(\sum_{i=0}^{2}a_{i}E_{\bar{\nu}_{e}}^{i}\bigg) ,
\end{equation}

\noindent where the $a_i$ parameters are specific to each isotope. Table \hyperlink{table1}{I} displays the fitted $a_i$ values for the four most dominant fissioning isotopes ($>$ 99\% of all fission) in nuclear reactors: \textsuperscript{235}U, \textsuperscript{238}U, \textsuperscript{239}Pu, and \textsuperscript{241}Pu. 

Though Eq.\,(\hyperlink{eq4}{4}) and Table \hyperlink{table1}{I} were determined for reactor antineutrinos relevant to IBD interactions ($>$ 1.8 MeV), it was assumed that Eq.\,(\hyperlink{eq4}{4}) was valid below this threshold. Our analysis (Section \ref{sec:analysis}) focuses on the high energy domain where directionality is strongest, therefore the assumption of extending the reactor spectrum below the IBD threshold is justified as it will not have any significant effects on our results. Furthermore, we neglect that the electron scattering cross-section has been shown to be $\sim$$1.5\sigma$ larger than predicted by the Standard Model at very low energies \cite{Vogel_Engel, Amsler}.

\hypertarget{table1}{}
\begin{table}[!htb]
\centering
\caption{Parameter values for Eq.\,(\protect\hyperlink{eq4}{4}). The values reported for \textsuperscript{235}U, \textsuperscript{239}Pu, and \textsuperscript{241}Pu are for thermal neutrons and the value for \textsuperscript{238}U is for 0.5 MeV neutrons \cite{Vogel_Engel}.}
\begin{tabular*}{\columnwidth}{ c  @{\extracolsep{\fill}} c  c  c }
\hline
{Isotope} & {$a_{0}$} & {$a_{1}$} & {$a_{2}$} \\
\hline
\textsuperscript{235}U & 0.870 & -0.160 & -0.0910 \\ 
\textsuperscript{238}U & 0.976 & -0.162 & -0.0790 \\ 
\textsuperscript{239}Pu & 0.896 & -0.239 & -0.0981 \\
\textsuperscript{241}Pu & 0.793 & -0.080 & -0.1085 \\
\hline 
\end{tabular*}
\end{table} 

The isotopic fissioning concentrations in a nuclear reactor will depend on the reactor design as well as the level of fuel burnup. In this work, fission concentrations of a typical mid-cycle pressurized light water reactor (PWR) were used (49.6\% \textsuperscript{235}U, 35.1\% \textsuperscript{239}Pu, 8.7\% \textsuperscript{238}U, and 6.6\% \textsuperscript{241}Pu) \cite{Zacek}. The emitted antineutrino energy spectra per fission for each isotope as well as the summation of the four isotopes weighted by the typical PWR concentrations are plotted in Fig.\,\hyperlink{fig2}{2} with dashed curves. As was mentioned before, reactor antineutrinos possess relatively low energies, with an average energy of about 1.5 MeV. Folding the incident antineutrino energy spectrum with the ES cross-section results in the observable/detectable spectrum shape in a detector. The detectable spectra per fission of the four isotopes as well as their weighted sum are plotted in Fig.\,\hyperlink{fig2}{2} with solid curves. The average detectable reactor antineutrino energy is approximately 2.5 MeV.

\begin{figure}[!htb]
\hypertarget{fig2}{}
\centering
\includegraphics[width=238pt]{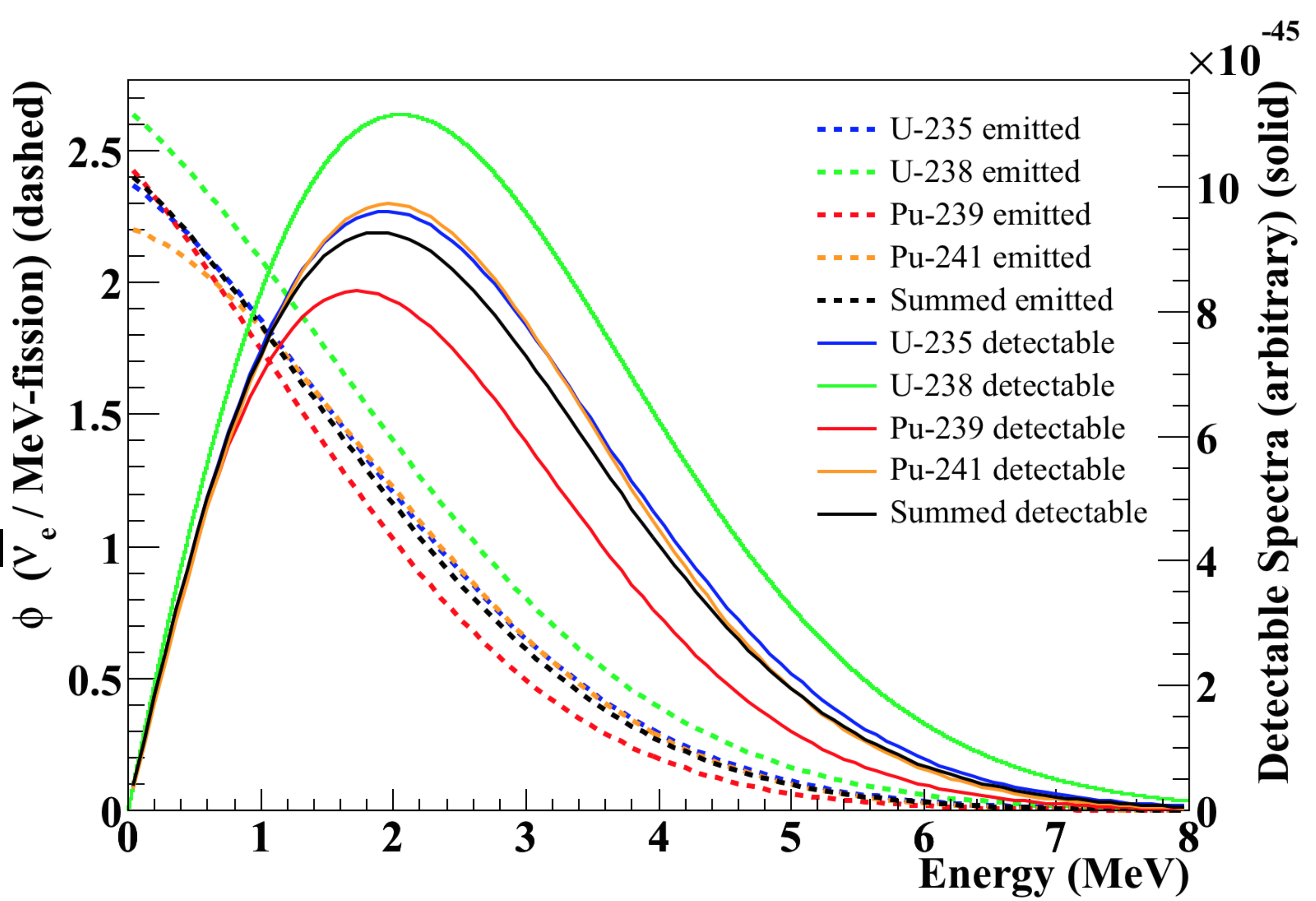}
\caption{Emitted (dashed) and detectable (solid) antineutrino energy spectra per fission from fission occurring in \textsuperscript{235}U, \textsuperscript{239}Pu, \textsuperscript{238}U, and \textsuperscript{241}Pu. The black lines represent the summation of the four isotopes weighted by the typical fission concentrations of a mid-cycle PWR (49.6\% \textsuperscript{235}U, 35.1\% \textsuperscript{239}Pu, 8.7\% \textsuperscript{238}U, and 6.6\% \textsuperscript{241}Pu).}
\end{figure}

\section{\label{sec:detectordesign}Proposed Detector Design} 

For this work we begin by considering a detector design based on the recently proposed WATCHMAN project \cite{Bernstein} - a kiloton scale WCD constructed from a large cylindrical stainless steel tank [see Fig.\,\hyperlink{fig3}{3(a)}]. The diameter and height of the cylinder are 15.8 m with a total water volume of about 3.1 kilotons. Photomultipler tubes (PMTs) are housed in a cylindrical structure 13.8 m in diameter, separating the detector into two distinct regions. The outer region serves as a veto for cosmic muons and the inner region as the target. There is approximately 2.1 kilotons of Gd-doped water in the target and 1 kiloton in the veto. The PMT support structure houses approximately 4300 30.48-cm (12-inch) Hamamatsu PMTs facing the target, with photocathode coverage near $40\%$, and 480 PMTs facing the veto. Within the target, a cylindrical fiducial volume (FV) was initially defined with a diameter and height of 10.82 m ($\sim$1 kiloton). The 1.5 m thick space between the PMT support structure and fiducial volume acts as a buffer region to enable better reduction of backgrounds from the PMTs and external radiation.

Like the WATCHMAN detector, the model assumes a single-core 3.758-GWt light water nuclear reactor located 13 km away. To model detector response, a Geant4 \cite{Agostinelli} based simulation package named Reactor Monitoring Simulation (RMSim) was used. RMSim is a modified version of WCSim \cite{WCSim}, a Geant4-based program for developing and simulating large WCDs. RMSim contains all relevant physics processes such as particle generation and transport, Cherenkov physics, optical photon production and transport, PMT sensitivity, digitization, and timing. Detailed detector geometry, materials, and optical properties for the WATCHMAN detector are also included. See Fig.\,\hyperlink{fig3}{3(b)} for a visualization of an antineutrino-electron scattering event in the simulated detector. Event reconstruction was handled by the fitter software code named BONSAI \cite{Smy}, originally developed for the Super Kamiokande (Super-K) experiment. We note that at the time of this writing, BONSAI has not been optimally tuned to the specifications of the proposed detector in the same way as for Super-K.

Note that the WATCHMAN detector was not originally designed with ES directional sensitivity in mind. The design was used here as a baseline model simply because a detailed Geant4-based simulation already existed. Therefore, this work considers modifications to certain features of the detector, namely the size and overburden, that will greatly improve ES directional sensitivity to the reactor antineutrino flux.

\begin{figure}[!htb]
\hypertarget{fig3}{}
\centering
\begin{tabular}{c c}
\includegraphics[height=125pt]{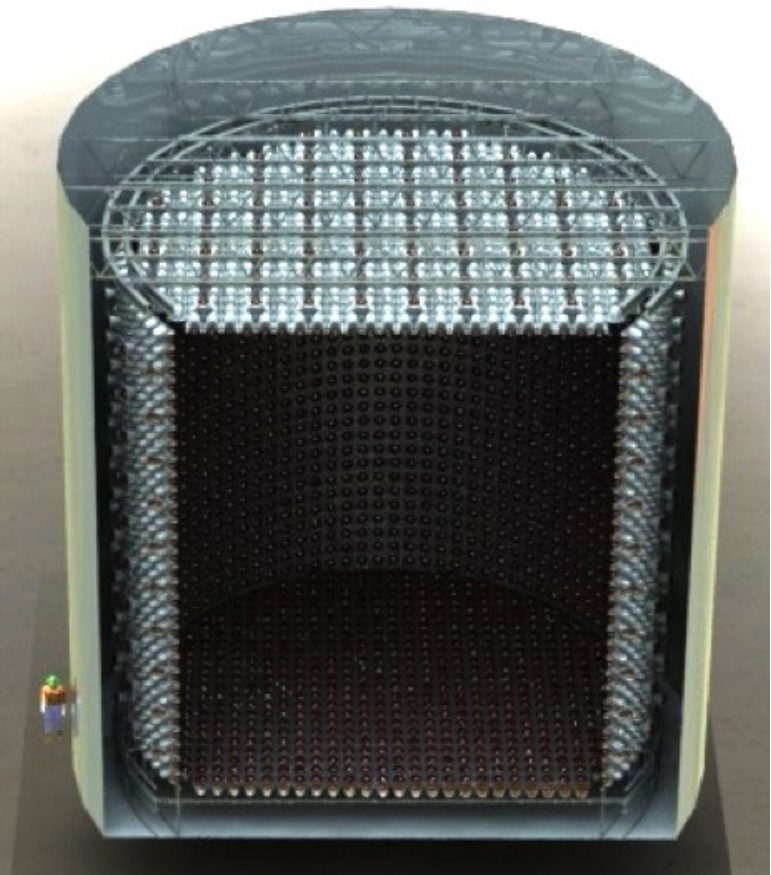} &
\includegraphics[height=125pt]{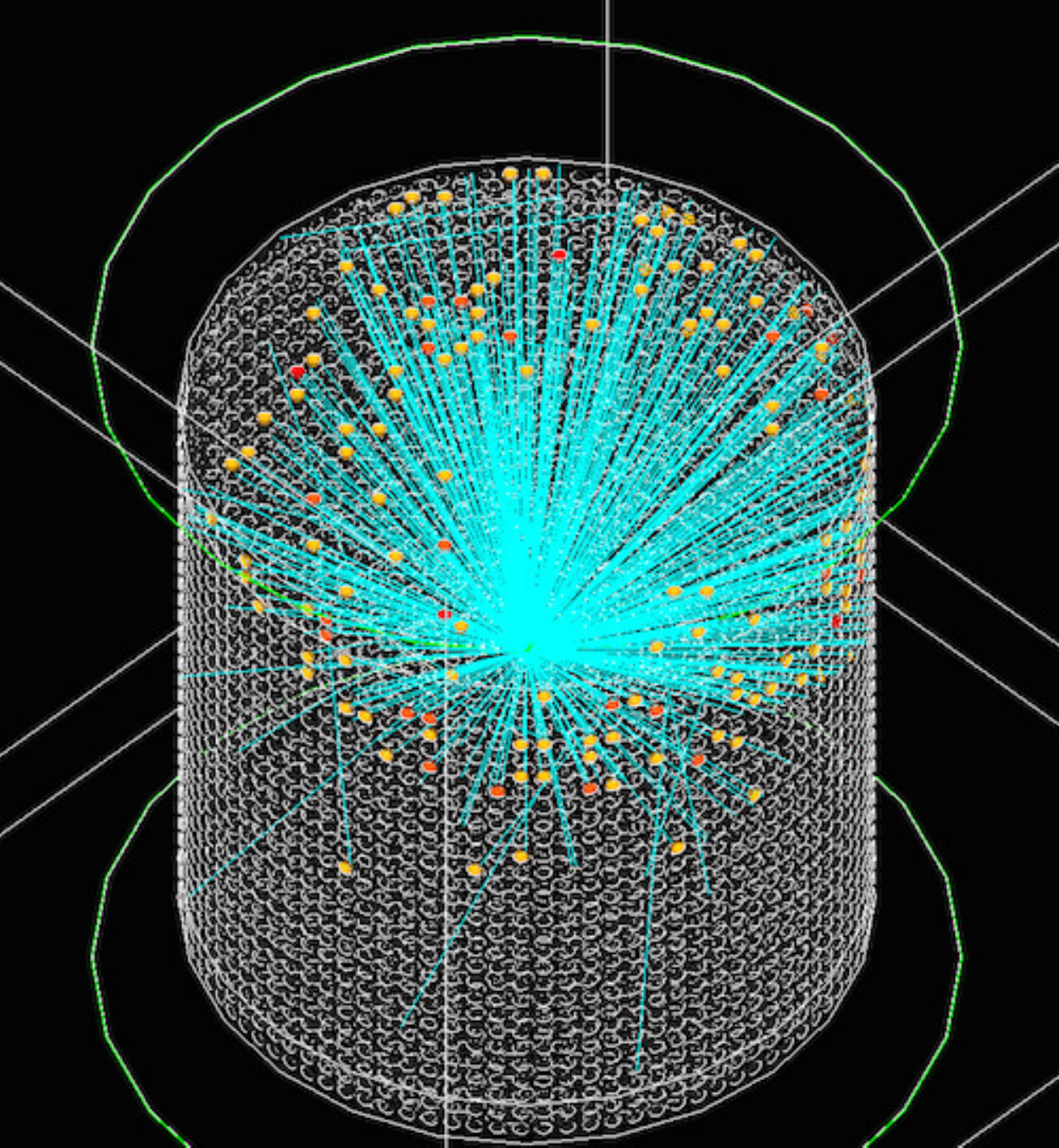} \\
\footnotesize (a) & \footnotesize (b)
\end{tabular}
\caption{(a) Basic design of the proposed kiloton WCD \cite{Bernstein}. (b) Visualization of an ES event in the proposed detector modeled in RMSim. The blue lines represent the Cherenkov light and the colored dots represent triggered PMTs.}
\end{figure}

\section{\label{sec:signal} Signal}

Neglecting oscillations, the reactor-based elastic antineutrino - electron scattering rate in a detector can be determined using 
\hypertarget{eq5}{}
\begin{equation} \tag{5}
R_{\bar{\nu}_{e}/e^{-}}=\frac{N_{e}}{4\pi D^{2}}\sum_i f_i\int \phi_i(E_{\bar{\nu}_{e}})   \sigma(E_{\bar{\nu}_{e}})dE_{\bar{\nu}_{e}}\,,
\end{equation}

\noindent where $N_e$ is the number of available target electrons, $D$ is the reactor-detector distance (cm), $f_i$ is the fission rate for the particular isotope $i$ (Hz), $\phi_i(E_{\bar{\nu}_{e}})$ is the number of antineutrinos produced per fission per MeV for isotope $i$ [see Eq.\,(\hyperlink{eq4}{4})], and $\sigma(E_{\bar{\nu}_{e}})$ is the energy dependent scattering cross-section (cm\textsuperscript{2}) [see Eq.\,(\hyperlink{eq1}{1})]. The sum runs over the four dominant fissionable isotopes in nuclear reactors mentioned in Table \hyperlink{table1}{I}, and the integral runs from 0 to 8 MeV, as in Fig.\,\hyperlink{fig2}{2}. Carrying out the calculation with the specifications outlined above results in about 9270 total scattering events in the kiloton FV over 5 years (not yet including detector response).

An elastic electron scattering generator was developed for RMSim to simulate the scattered electrons. The generator calculates the total number of expected interactions for any desired detector size, acquisition time, standoff distance, reactor power level, and fission isotopics using Eq.\,(\hyperlink{eq5}{5}). It then generates a sample of scattering events by sampling position, energy, and direction using Eqs.\,(\hyperlink{eq2}{2}-\hyperlink{eq4}{4}).

Five years worth of ES events were simulated in RMSim and reconstructed using the BONSAI fitter software. The reconstructed cosine of the scattering angles are shown in Fig.\,\hyperlink{fig4}{4(a)} with a value of $\cos\theta = 1$ denoting a complete forward scatter of the electron. The reconstructed distribution appears to follow an exponential-like distribution peaking at $\cos\theta = 1$. RMSim imposes a triggering threshold of 16 photoelectrons, and it can be seen from the plot that only 1550 ($\sim$17\%) of the original 9270 ES events trigger the detector. Figure \hyperlink{fig4}{4(b)} shows the detector response of the triggered ES events in terms of detected photoelectrons. The distribution follows a decreasing exponential that extends to $\sim$140 photoelectrons. We note that we expect approximately 10 photoelectrons/MeV in the detector at low energies. While the PMT coverage is similar (40\%), this is slightly larger than the Super-K results of 6 photoelectrons/MeV \cite{Renshaw} because of the higher quantum efficiency of the PMTs used here.

\begin{figure}[!htb]
\hypertarget{fig4}{}
\centering
\begin{tabular}{c}
\includegraphics[width=238pt]{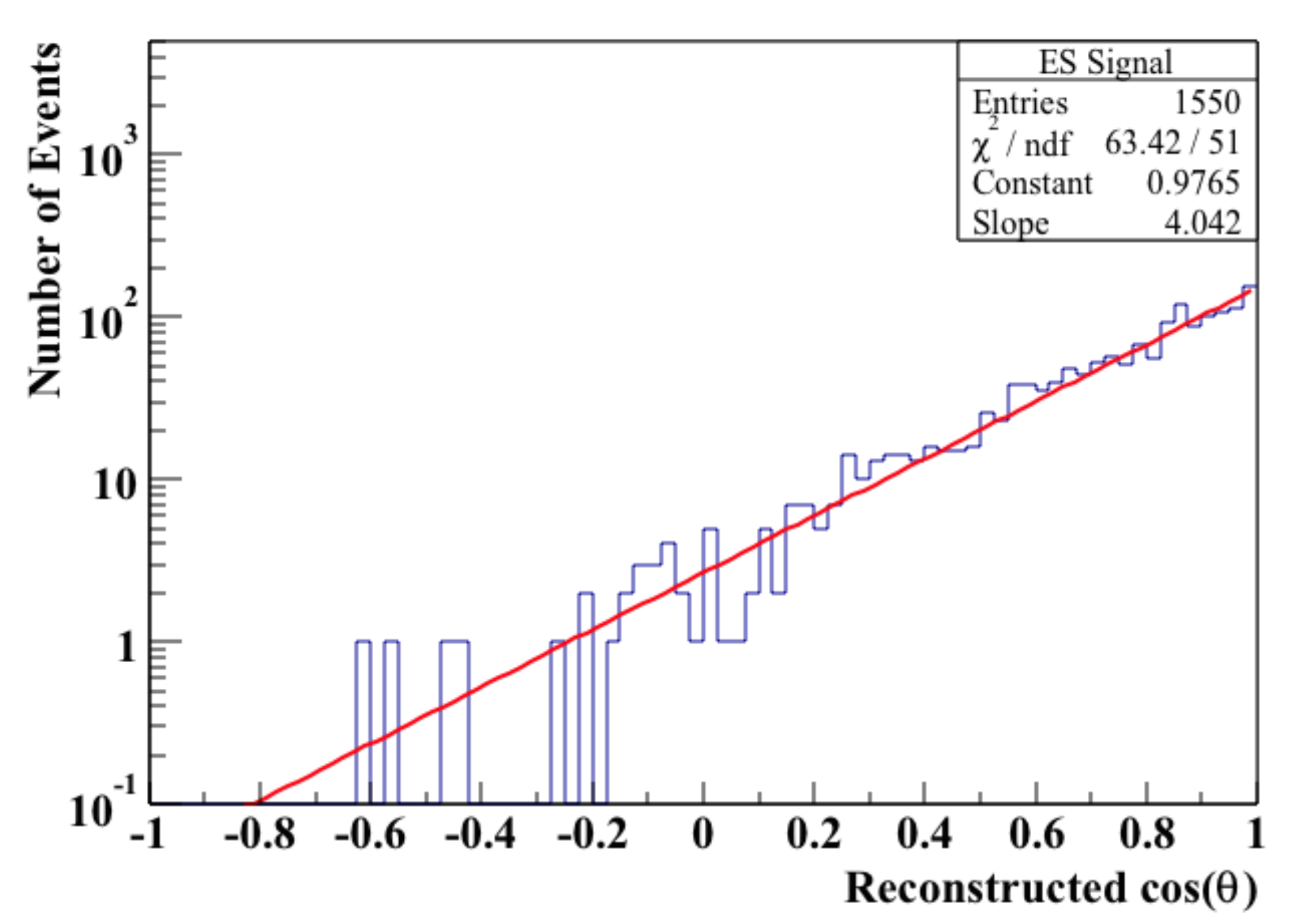} \\
\footnotesize (a) \\
\includegraphics[width=238pt]{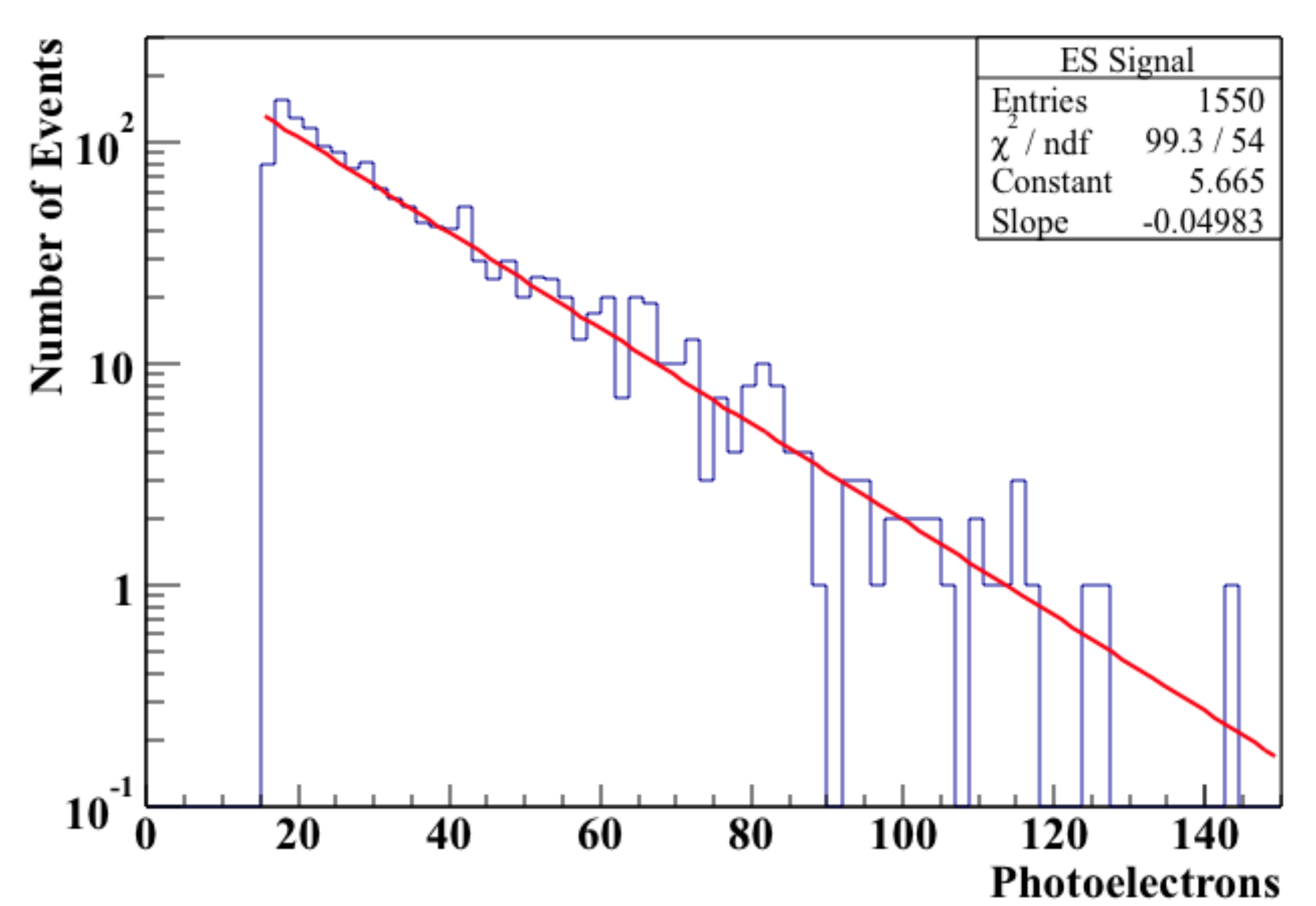} \\
\footnotesize (b)
\end{tabular}
\caption{(a) Reconstruction of the cosine of the scattering angle distribution for 5 years of reactor ES in the proposed WCD. (b) Detected photoelectron distribution of the 5-year triggered signal.}
\end{figure}

\section{\label{sec:background}Backgrounds}

Due to the low count rates associated with antineutrino detection, the background levels in the detector must be kept to a minimum to maintain suitable statistics. Several potential sources of background exist for ES including cosmogenic radionuclides, high-energy gamma rays from the steel vessel and the rock surrounding the detector, solar neutrinos, misidentified IBD events from the reactor, PMT gamma rays, and water-borne radon. All were assumed to be distributed isotropically in direction (neglecting the obvious anisotropy of solar neutrinos). Therefore, in a directional cosine plot, the reactor ES signal should appear as a peak in the forward direction atop a flat background.

\subsection{\label{sec:radionuclides}Cosmogenic radionuclides}

Cosmic muons and the hadronic showers they produce can interact with the oxygen atoms in the target region water to create long-lived ($>$ 1 s) radionuclides. If they beta decay in the inner detector region, the resultant electrons can trigger the PMTs and mimic the ES signal. The cosmogenic radionuclide production yields at Super-K have been estimated using \textsc{fluka} \cite{Li_Beacom}. Recently, measurements of the production yields in Super-K were also published \cite{Zhang}. Table \hyperlink{table2}{II} shows the theoretical and measured results for the five isotopes determined to be the most relevant for reactor antineutrino-electron scattering due to their long lifetimes and/or high yields. The theoretical yields were used to determine the production rates in the case of \textsuperscript{15}C, \textsuperscript{11}Be, \textsuperscript{8}B, and \textsuperscript{8}Li because numerical values were provided for each isotope (the theoretical yields for \textsuperscript{8}B and \textsuperscript{8}Li also provide a conservative estimate over the measured values). In the case of \textsuperscript{16}N, the measured yield (the larger of the two) was used in the production rate calculation.

\hypertarget{table2}{}
\begin{table*}[!htb]
\setlength{\tabcolsep}{7pt}
\centering
\caption{Cosmogenic radionuclide production yields in water calculated by \cite{Li_Beacom} and measured by \cite{Zhang} for the Super-K detector. Only the isotopes determined to be relevant to reactor antineutrino-electron scattering are considered.}
\begin{tabular}{ c c c c c c  }
\hline
\multirow{2}{*} {Isotope} & 
\multirow{2}{*} {\begin{tabular}{c} Half-life\\ (s)\end{tabular}} & 
\multirow{2}{*} {\begin{tabular}{c} Decay\\ Mode\end{tabular}} &
\multirow{2}{*} {\begin{tabular}{c} Theoretical Yield\\ (10$^{-7}\mu^{-1}\text{g}^{-1}\text{cm}^{2}$)\end{tabular}} & 
\multirow{2}{*} {\begin{tabular}{c} Measured Yield\\ (10$^{-7}\mu^{-1}\text{g}^{-1}\text{cm}^{2}$)\end{tabular}} & 
\multirow{2}{*} {\begin{tabular}{c} Primary Process\\ (on \textsuperscript{16}O) \end{tabular}} \\
&&&&& \\ 
\hline 
\textsuperscript{16}N & 7.13 & $\beta^{-}\gamma \,(66\%)$, $\beta^{-} \,(34\%)$ & \hspace{-1em}18 & 23.4 $\pm$ 1.9 $\pm$ 1.7 & $(n, p)$ \\ 
\textsuperscript{15}C & 2.45 & $\beta^{-}\gamma \,(63\%)$, $\beta^{-} \,(37\%)$ & 0.8 & $<$3.9 & $(n, 2p)$ \\ 
\textsuperscript{11}Be & \hspace{-1em}13.8 & $\beta^{-} \,(55\%)$, $\beta^{-}\gamma \,(45\%)$ & 0.8 & $<$10.0 & $(n, \alpha + 2p)$ \\ 
\textsuperscript{8}B & 0.77 & $\beta^{+}$ & 5.8 & \multirow{2}{*}{4.9 $\pm$ 0.2 $\pm$ 0.2} & $(\pi^{+}, \alpha + 2p + 2n)$ \\ 
\textsuperscript{8}Li & 0.84 & $\beta^{-}$ & \hspace{-1em}13 && $(\pi^{-}, \alpha$ $ + $\textsuperscript{ 2}H $ +$ $ p +n)$ \\
\hline 
\end{tabular}
\end{table*}

The production yields of Table \hyperlink{table2}{II} can be converted to production rates using
\hypertarget{eq6}{}
\begin{equation} \tag{6}
R_{i}=\rho Y_{i}L_{\mu}R_{\mu}\,,
\end{equation}

\noindent where $\rho$ is the density of the target (g\,cm\textsuperscript{-3}), $Y_i$  is the yield of isotope \emph{i} (10\textsuperscript{-7}\,$\mu$\textsuperscript{-1}\,g\textsuperscript{-1}\,cm\textsuperscript{2}), $L_\mu$ is the average muon path length in the detector (cm), and $R_\mu$ is the muon rate (Hz). To determine how the radionuclide backgrounds scale with depth, we began by assuming a water detector at the Kamioka Liquid scintillator ANtineutrino Detector (KamLAND) experiment location, which is at the same depth as Super-K. Given the published showering and non-showering muon rates at KamLAND (0.037 Hz and 0.163 Hz, respectively \cite{Abe1}) and the proportion of radionuclides produced by the showering component at this depth is $70\%$ \cite{Abe2}, we can use the predictions of Table \hyperlink{table2}{II} for water to estimate the radionuclide production rates per unit volume at any depth by scaling with the showering and non-showering muon rates. The total muon rate scaling was obtained from the analytical expression for the differential muon intensity (cm$^{-2}$ s$^{-1}$) in the flat-earth approximation provided by Mei and Hime:
\hypertarget{eq7}{}
\begin{equation}  \tag{7}
I_\mu(h_0)=(67.97e^{\frac{-h_0}{0.285}}+2.071e^{\frac{-h_0}{0.698}})\times10^{-6}\,,
\end{equation}

\noindent where $h_0$ is the vertical depth (km.w.e.) \cite{MeiHime}. Similarly, we employed their expression for the muon energy spectrum for any slant depth (the averaged distance traveled through rock by muons at an experiment) $h$ (km.w.e.):
\hypertarget{eq8}{}
\begin{equation}  \tag{8}
\frac{dN}{dE_\mu}=Ae^{-bh(\gamma_\mu -1)} \big[E_\mu+ \epsilon_\mu(1 - e^{-bh})]^{-\gamma_\mu}\,,
\end{equation}

\noindent where $A$ is a normalization constant with respect to the differential muon intensity at a particular depth, $b$ = 0.4 km.w.e.$^{-1}$, $\gamma_\mu$ = 3.77, $\epsilon_\mu$ = 693 GeV, and $E_\mu$ is the muon energy in GeV. Once a muon spectrum is generated for an assumed depth, we can calculate the average muon energy. Previously published estimates of the mean muon energy at KamLAND have ranged from 198 GeV to 285 GeV \cite{Hagner, MeiHime, GalbiatiBeacom}. A depth of 2350 m.w.e.\,\,produces an average muon energy consistent with the midpoint of the range (240 GeV), and was therefore accepted as our best estimate of the slant depth of KamLAND. We also make a simplifying assumption that there is an energy above which all muons form showers, which we define as the  ``showering equivalent energy". For KamLAND, where $18\%$ of the muon flux is showering, the ``showering equivalent energy'' is 354 GeV. Using the same approach and by matching the total muon flux reported in \cite{IMB}, the depth of the Irvine-Michigan-Brookhaven (IMB) detector site (the same site for the proposed WATCHMAN detector) was estimated to be 1540 m.w.e. This result is close to the 1570-m.w.e.\,\,depth reported by IMB \cite{IMB}.

The outer veto is used to identify and reject spallation events following muons entering the detector. For this work, an additional muon veto must be applied to reduce cosmogenic radionuclide decays. Following a muon that traverses the inner FV region, all subsequent events within 2 m of a showering muon track, or 1 m of a non-showering muon track are removed for a period of time dependent upon depth. Veto time adjustments as a function of depth are described in further detail in Section \ref{sec:depth}. The detector live time at each depth was calculated conservatively assuming that all muons traverse the entire length of the cylindrical FV.

Applying the tubular veto above, the rate of each of the five major radionuclide components were calculated as a function of depth. Due to its long lifetime and large yield, \textsuperscript{16}N significantly dominates the mix, producing $\sim$90\% of the total. Uncertainties in the vertex reconstruction, which result in some radionuclide events being reconstructed outside the tubular veto regions surrounding the muon tracks, were also determined via independent simulations and included in the calculations.

\subsection{\label{sec:pmts}PMT gamma rays}

The PMT glass will contain trace amounts of natural U, Th, and K. The decays of \textsuperscript{208}Tl  (from the Th decay chain) and \textsuperscript{40}K will produce 2.6-MeV and 1.4-MeV gamma rays, respectively. Most of these will interact outside the FV, but due to the uncertainty in the event reconstruction, some events will be reconstructed inside, contributing to the background. An arbitrary number of PMT gamma rays were simulated in RMSim and the black curve (right diagonal shading) in Fig.\,\hyperlink{fig5}{5} shows the distance from the reconstructed interaction vertex to the nearest PMT for each event. From the figure, it is clear that a significant number of events are reconstructed inside the FV ($>$ 150 cm away from the PMTs), forming two distinct groups. Near the PMTs, the black curve appears to follow an exponential, whereas further away from the PMTs an almost flat distribution is observed. To improve upon the results, we attempt to remove the poorly fit events. By applying a cut to the log likelihood fit parameter ($\geq25$) and the number of triggered PMTs ($\geq25$), roughly half of the events are removed, leaving an exponential distribution with respect to the distance to the PMTs (shown in blue and left diagonal shading in Fig.\,\hyperlink{fig5}{5}). 

\hypertarget{fig5}{}
\begin{figure}[!htb]
\centering
\includegraphics[width=222pt]{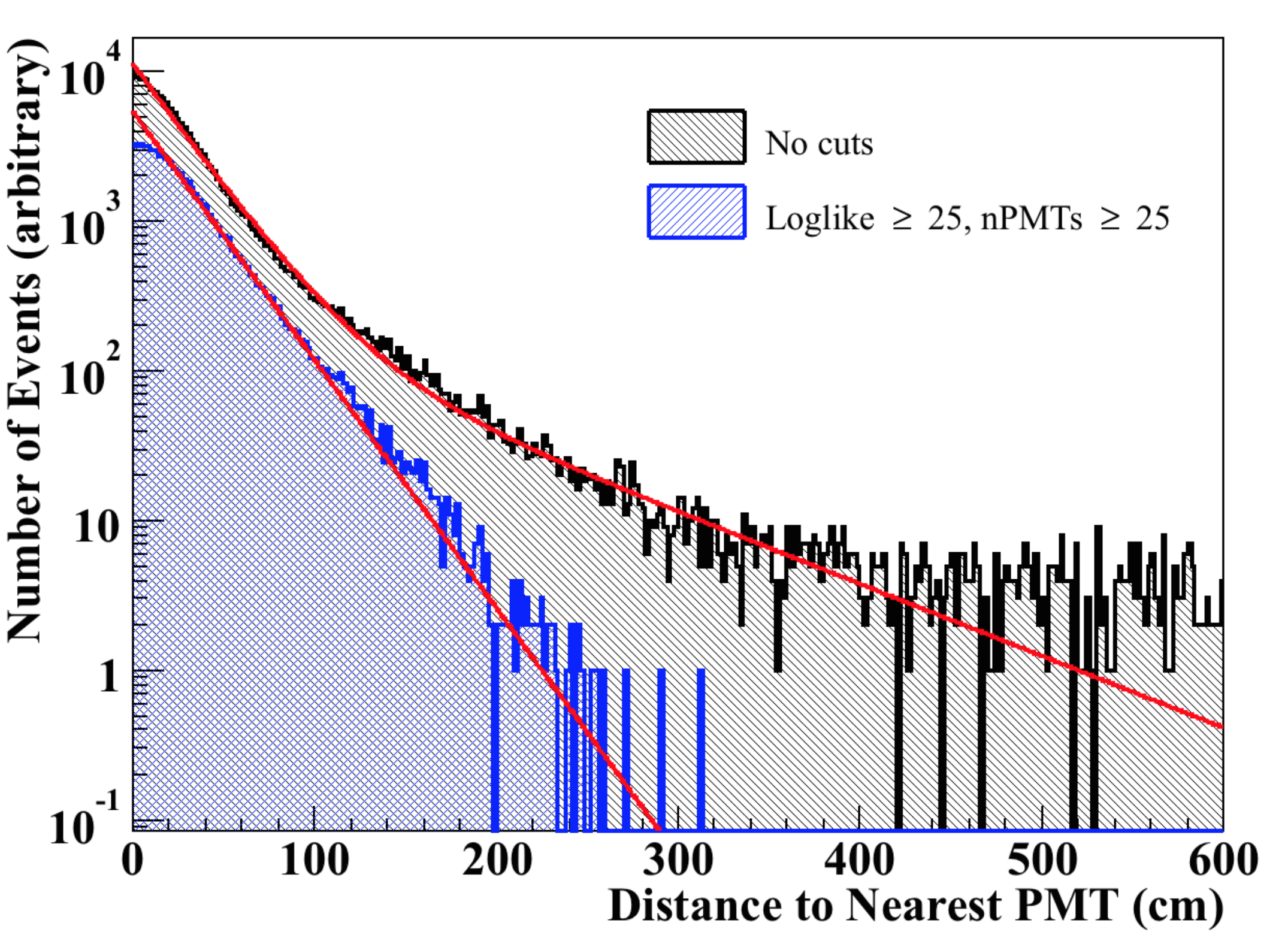}
\caption{PMT-based background events as a function of the distance from the reconstructed vertex to the nearest PMT (black and right diagonal shading). The blue curve (left diagonal shading), which requires both the triggered PMT count (nPMTs) and the log likelihood (Loglike) to be $\geq$ 25, follows an exponential distribution.}
\end{figure}

The exponential behavior of the blue curve (left diagonal shading) in Fig.\,\hyperlink{fig5}{5} is a promising result, if realizable in practice. It implies that the PMT gamma ray background can be reduced to a subdominant level with a large enough buffer region. To reduce the PMT gamma ray backgrounds with a fixed detector size however, the FV must be decreased to allow for a sufficient buffer thickness. This will result in a significant reduction in the number of detectable ES interactions. Assuming an exponential distribution with respect to distance from the PMTs, the PMT gamma ray background can be estimated for any sized FV using the assumed impurity levels of Th and \textsuperscript{40}K in the glass. In this work, the PMTs are assumed to have similar radioactivity levels as the low-background 25.4-cm (10-inch) Hamamatsu PMTs employed at the Double CHOOZ detector with Th and \textsuperscript{40}K impurity concentrations of 0.03 ppm and 20 ppm, respectively \cite{Felde}.

\subsection{\label{sec:radon}Water-borne \texorpdfstring{\textsuperscript{222}}{222}Rn/\texorpdfstring{\textsuperscript{214}}{214}Bi}

The beta decay of \textsuperscript{214}B (a daughter product of the \textsuperscript{238}U decay chain product \textsuperscript{222}Rn, Q = 3.3 MeV) in the target region will also contribute a significant amount of background to the ES signal. The presence of \textsuperscript{222}Rn/\textsuperscript{214}Bi in the water can occur due to a variety of processes. Some may result due to trace amounts of naturally occurring \textsuperscript{238}U present in the water, dissolved \textsuperscript{222}Rn that has migrated out of the PMT glass, and from radon gas entering the detector from the mine air. The Sudbury Neutrino Observatory (SNO) heavy water neutrino detector has reported an inner detector radon contamination of 10\textsuperscript{-14} gU/gD$_2$O, assuming the U is in secular equilibrium with \textsuperscript{222}Rn \cite{Belvis}. Assuming this level of contamination in the proposed light water detector results in about about 10\textsuperscript{4} \textsuperscript{214}Bi decays per day somewhere in the 1000-m\textsuperscript{3} FV, of which approximately 20\% survive the Geant4 detector simulation trigger condition (16 photoelectrons).

Actual radon levels achievable in a real detector will rely on the water recirculation methods employed, as well as the radon concentration in the mine air, both of which could be significantly different than SNO. The SNO detector also employs an acrylic barrier between the heavy water target and the light water buffer. The acrylic, while it impedes the migration of radon from the PMTs to the target, might also be a mild source of radon. One might envision a different water flow scheme, in which radon free fresh water is injected inside the target and directed outward via positive pressure, could achieve reductions in radon contamination relative to SNO. In this work, since it is difficult to predict physically achievable radon concentrations, we simply assume similar concentrations to SNO as well as hypothetical situations in which the radon contamination can be reduced further.

\subsection{\label{sec:otherbackgrounds}Other backgrounds}

The backgrounds due to gamma rays from the detector steel vessel and the surrounding rock were determined using a study performed by the Isotope Decay At Rest (IsoDAR) collaboration on the KamLAND detector \cite{Toups}. IsoDAR assumed a 5-m sphere FV at KamLAND, thus the results from \cite{Toups} were scaled to account for the much larger cylindrical FV of the proposed detector (1000 m\textsuperscript{3}). Specifically, the estimates were scaled using the difference in the fiducial surface areas. This method assumes the proposed detector steel vessel will have similar cleanliness levels as KamLAND and the surrounding rock will be of similar composition to the KamLAND mine. The differences in densities and gamma attenuation lengths between the scintillator used in KamLAND and the water used in the detector under study, as well as the differences in gamma path lengths for the spherical and cylindrical geometries were neglected. All gamma rays that reached the FV were assumed to interact.

The \textsuperscript{8}B solar neutrino background was also determined by scaling from \cite{Toups}. Assuming the neutrino flux is constant with depth, the interaction rate is dependent solely on the number of available targets, which is proportional to the fiducial mass. Therefore the solar neutrino background estimation in \cite{Toups} was scaled according to the difference in the the KamLAND fiducial mass (0.408 kilotons) and the proposed detector fiducial mass (1 kiloton). 

The scaled steel, rock, and solar neutrino results from \cite{Toups} were corrected for the difference in detector live time between KamLAND (56.2\%) and the model at any depth. Corrections were also included to account for the 3-MeV visible energy threshold used in \cite{Toups}.

If the neutron from a reactor-based IBD event is not detected within the time or spatial coincidence requirements, or it is simply not captured, then the lone positron signal will mimic ES. These misidentified IBD backgrounds were estimated assuming an IBD interaction rate of 20 events per day and a 20\% missed neutron rate as in \cite{Bernstein}.

\section{\label{sec:analysis}Analysis}

As mentioned in Section \ref{sec:background}, background events are assumed to be isotropic in direction. Reconstructed ES signal events exhibit an exponential behavior towards $\cos\theta = 1$. Therefore, in a plot of the cosine of the scattering angle, we expect the total signal to follow the behavior of a constant plus an exponential curve as in
\hypertarget{eq9}{}
\begin{equation} \tag{9}
y=A+Be^{Cx},
\end{equation}

\noindent where $A$, $B$, and $C$ are free parameters in the fit to the data. To determine the statistical significance of the ES signal, an arbitrarily large independent sample of ES events was simulated to determine the exponential slope parameter $C$. With the slope parameter fixed, the uncertainty in the exponential normalization parameter $B$ was used to determine the uncertainty and statistical significance of the signal. This analysis method would only be possible in practice if the exponential slope could be obtained \emph{a priori} using directional calibrations, such as the electron accelerator at Super-K \cite{Koshio}.

Figure \hyperlink{fig6}{6} displays the detector response (photoelectron production) from all sources of background except PMTs, as well as the ES signal for a time period of one year in a 3-kiloton water detector at the depth of the KamLAND detector (2350 m.w.e.). PMT backgrounds were not included because the rate normalization, which ranges from dominant to minor, depends entirely on the arbitrary fiducial volume chosen. From the plot it is clear that \textsuperscript{222}Rn/\textsuperscript{214}Bi dominates the total number of backgrounds, particularly at low energies. At higher energies and shallower depths, radionuclides begin to dominate.

\hypertarget{fig6}{}
\begin{figure}[!htb]
\centering
\includegraphics[width=255pt]{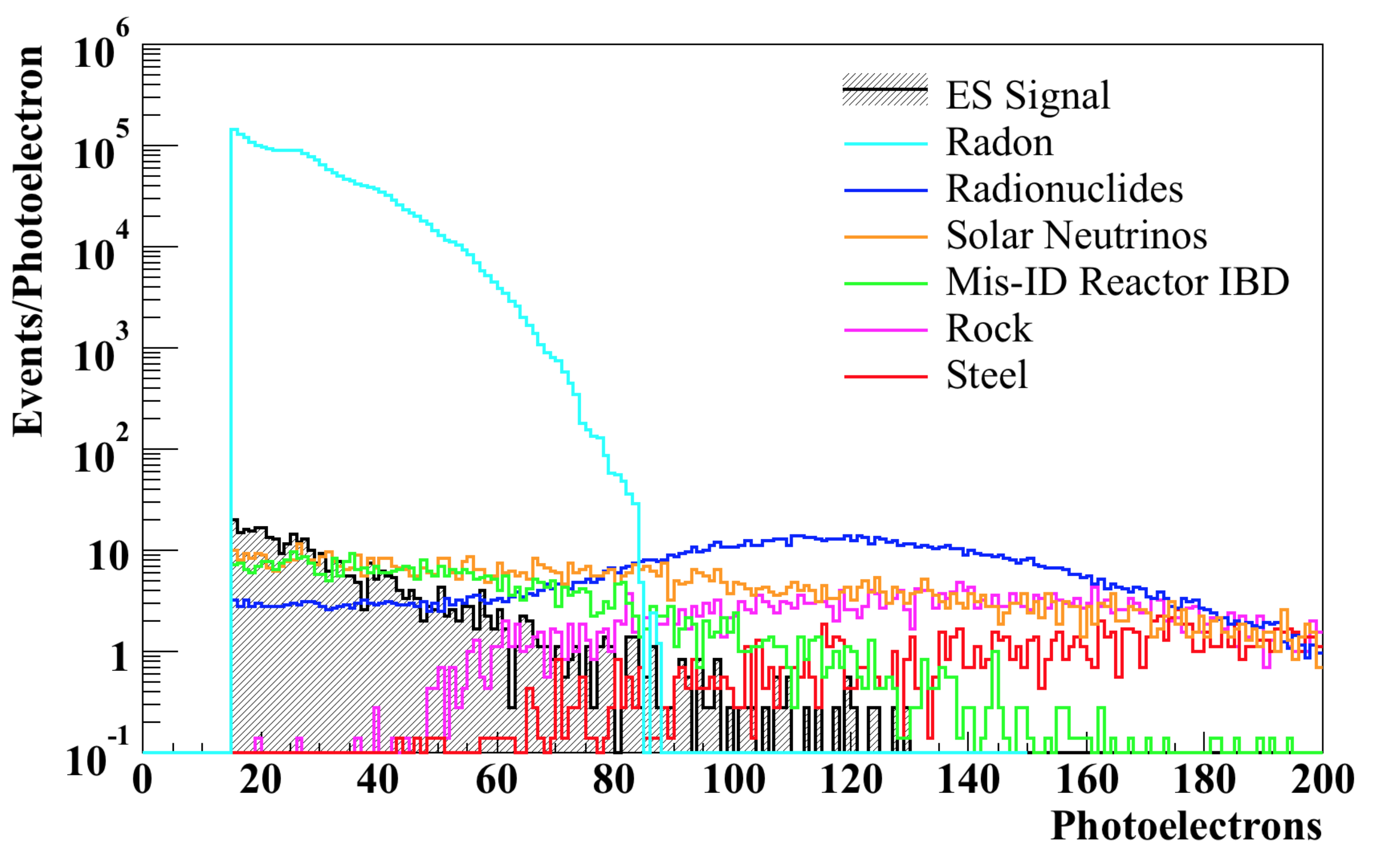}
\caption{The most significant backgrounds expected in a kiloton FV WATCHMAN-like detector over a one-year data acquisition period together with the ES signal at the same depth as KamLAND (2350 m.w.e.). Water-borne \textsuperscript{222}Rn/\textsuperscript{214}Bi and cosmogenic radionuclides represent the most important background types shown here. Note, sufficient distance between the PMTs and the fiducial volume was assumed to reduce PMT backgrounds to a subdominant level. In the following, we investigate the sensitivity of our model to many of these backgrounds as a function of energy, depth, and fiducial volume.}
\end{figure}

Based on the spectral shapes shown in Fig.\,\hyperlink{fig6}{6}, it is worth investigating if detector sensitivity has some dependence on the amount of detected energy. First, note that the ES shape extends to higher energies than the radon background. The radionuclide background, however, begins to dominate at high energy. Therefore, in the following section, detector sensitivities are presented for small slices in energy (25 $\rightarrow$ 65
and 60 $\rightarrow$ 90 triggered PMTs) and at different depths. At higher energies, the PMT based backgrounds are both lower in number and more accurately reconstructed, and thus larger FVs can be used. For the radon, it is clear that a significant improvement in contamination (relative to the SNO levels) would need to be made before ES directionality might be achievable. We cannot comment on whether a dedicated R\&D campaign or a new scheme of optimized water flow might be able to achieve significant improvements. Here we simply calculate the sensitivities that would result if significant reductions were achieved.

\subsection{\label{sec:depth}Sensitivity vs. depth}

We now investigate the overall behavior of the directional sensitivity as a function of depth using the showering and non-showering muon rate scalings with depth determined with the methods described in Section \ref{sec:radionuclides}. For the purpose of this work, we only consider depths from 1500 to 3000 m.w.e. The showering and non-showering scaling factors (relative to the KamLAND depth) are shown in Fig.\,\hyperlink{fig7}{7(a)}.

Using the muon scalings, the radionuclide background and detector live time were determined as a function of depth. Because the muon rate decreases significantly with depth, the position sensitive veto time can be increased to remove more radionuclide background without suffering any live time losses. Therefore, the tubular veto time was increased with depth to maintain a live time approximately equal to the KamLAND live time (56\%). A maximum veto time of 20 s was arbitrarily imposed since the radionuclides will migrate outside of the tubular veto if given enough time. Figure \hyperlink{fig7}{7(b)} displays the veto times and detector live times as a function of depth used in subsequent calculations. A veto time of 20 s is reached at 1900 m.w.e.\,\,and remains fixed at deeper depths.

\begin{figure}[!htb]
\hypertarget{fig7}{}
\centering
\begin{tabular}{c}
\includegraphics[width=220pt]{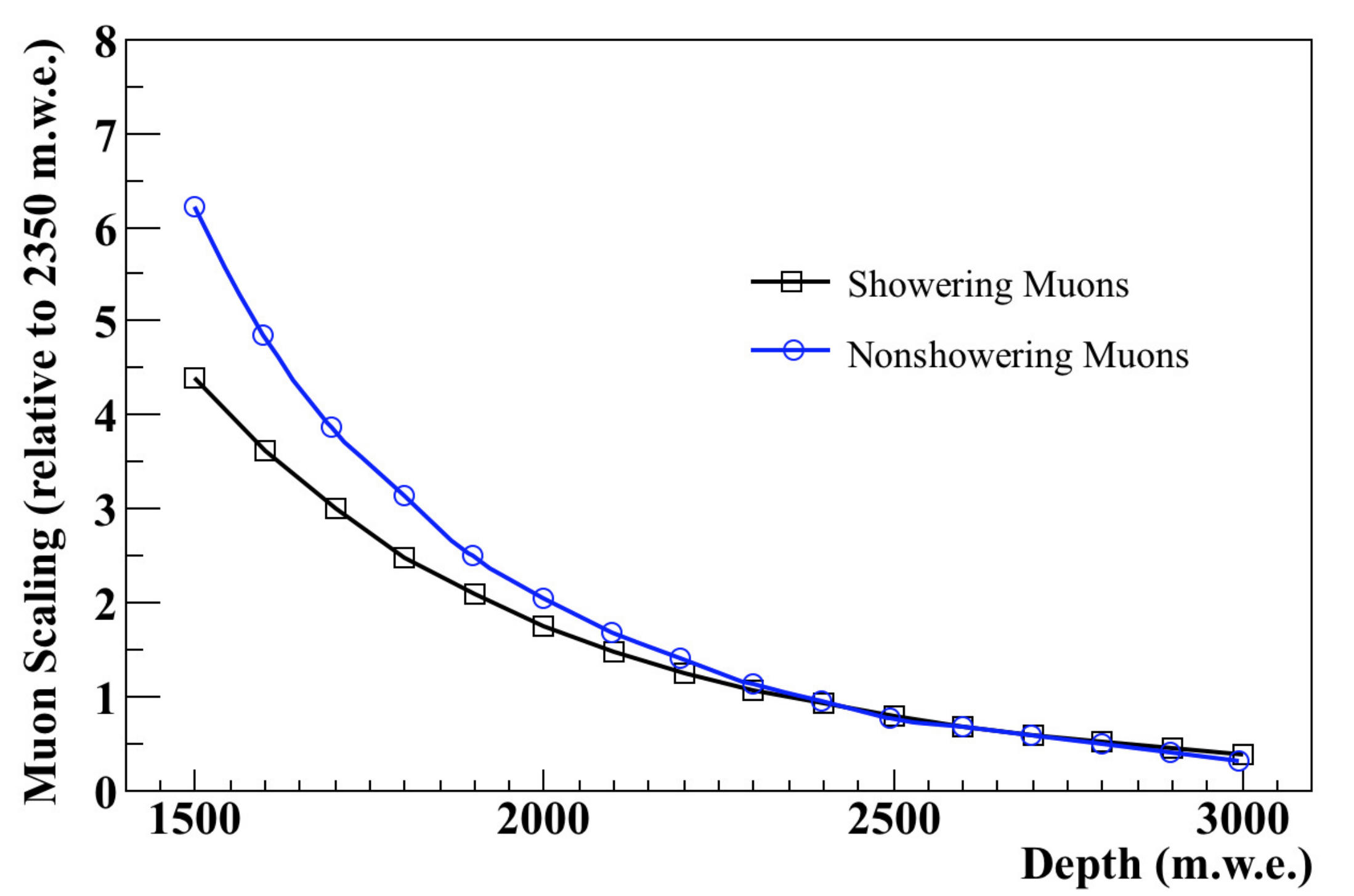} \\
\footnotesize (a) \\
\includegraphics[width=236pt]{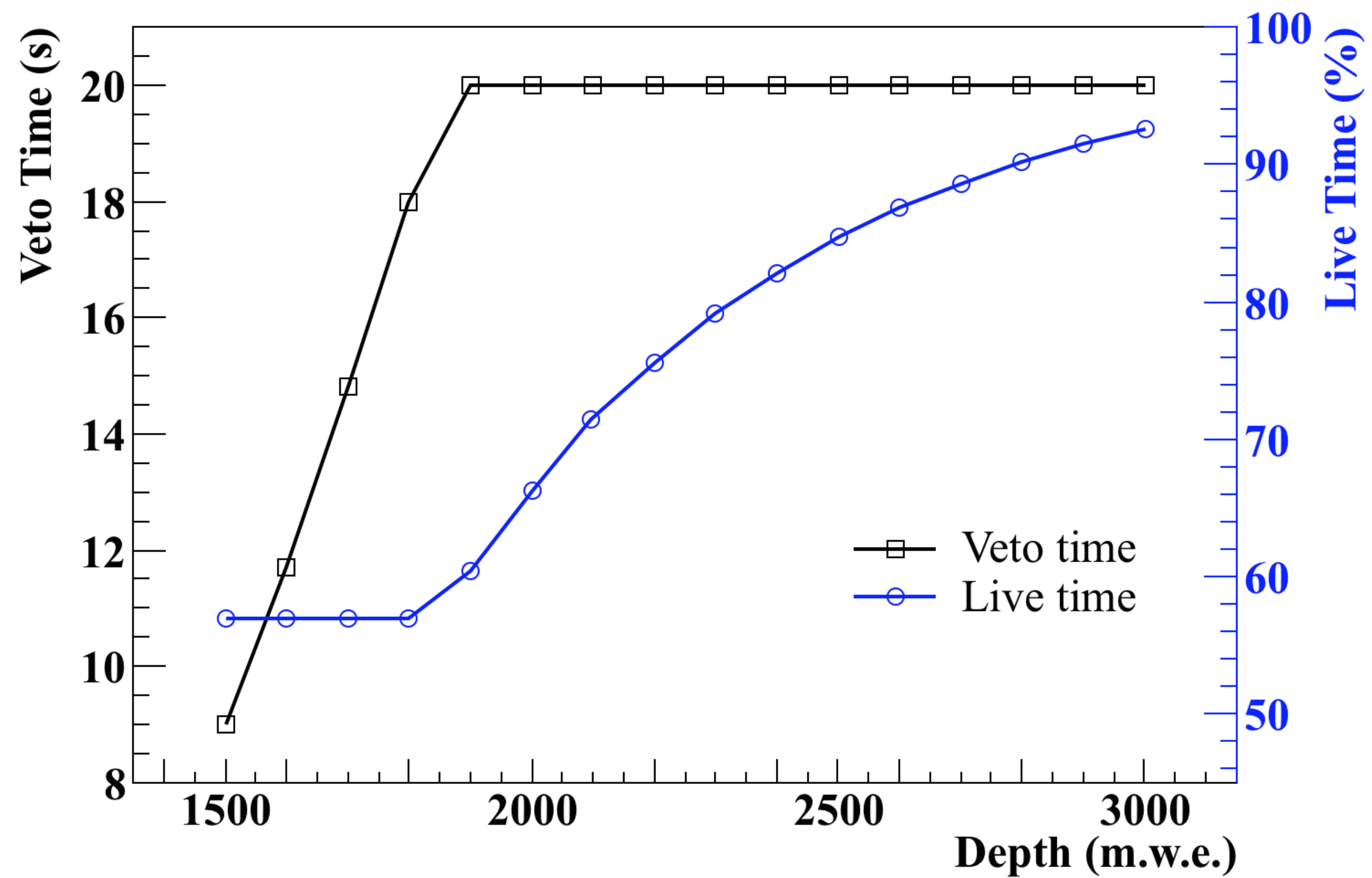} \\
 \footnotesize (b)
\end{tabular}
\caption{(a) Showering and non-showering muon rates as a function of depth (relative to KamLAND) determined from \cite{MeiHime}. (b) Veto times used in the position sensitive veto system as a function of depth and the resultant detector live times. The veto time was varied as a function of depth in order to retain a constant live time up to a maximum veto time of 20 s, which was reached at 1900 m.w.e.}
\end{figure}

The average statistical significances as a function of depth and radon contamination relative to SNO were then determined. The results are shown in Fig.\,\hyperlink{fig8}{8} for radon levels of 1 $\times$ SNO [Fig.\,\hyperlink{fig8}{8(a)}], 10\textsuperscript{-2} $\times$ SNO [Fig.\,\hyperlink{fig8}{8(b)}], and 10\textsuperscript{-4} $\times$ SNO [Fig.\,\hyperlink{fig8}{8(c)}]. As an example, \hyperlink{appendixA}{Appendix A} displays a detailed breakdown of the expected number of elastic scattering signal and background events in the two different energy ranges for a kiloton sized WATCHMAN-like detector at the same depth as KamLAND (2350 m.w.e.). Repeated multiple independent data samples were used to calculate the mean significance per 5-year experiment. Error bars are included in Fig.\,\hyperlink{fig8}{8(a)-(c)} and represent the uncertainty in the mean of the many independent 5-year experiments, however are too small to be observed here. Note, the results of a single experiment will produce sensitivity values distributed around the mean with an uncertainty of approximately 1$\sigma$. With no reduction in radon (relative to SNO), directionality does not seem to be possible at any depth with a kiloton sized detector. If the radon contamination is significantly reduced (by four orders of magnitude), the 25 $\rightarrow$ 65 slice produces the most significant signal. This is clearly observed in Fig.\,\hyperlink{fig8}{8(c)}, where a 3$\sigma$ significance can be obtained using this slice starting at about 1900 m.w.e.

The total detector size (including the fiducial, buffer, and veto) required to obtain a significant (3$\sigma$) signal was also considered for the three radon levels in Fig.\,\hyperlink{fig8}{8(a)-(c)}. This was done assuming both the signal and background scale linearly with the FV, while significance scales with the signal ($S$) to background ($B$) ratio ($S/\sqrt{B}$). The respective buffer and veto thicknesses for each energy range were then added to determine the total detector size. The results are shown in Fig.\,\hyperlink{fig8}{8(d)-(f)} for all three radon levels. Error bars are included and represent the uncertainties in Fig.\,\hyperlink{fig8}{8(a)-(c)} propagated through the calculation. However, once again, the error bars represent the uncertainty in the mean and do not represent the uncertainty of a single experiment. 

\begin{figure*}[!htb]
\hypertarget{fig8}{}
\centering
\setlength{\tabcolsep}{3.pt}
\begin{tabular}{ccc}
\includegraphics[height=0.255\linewidth]{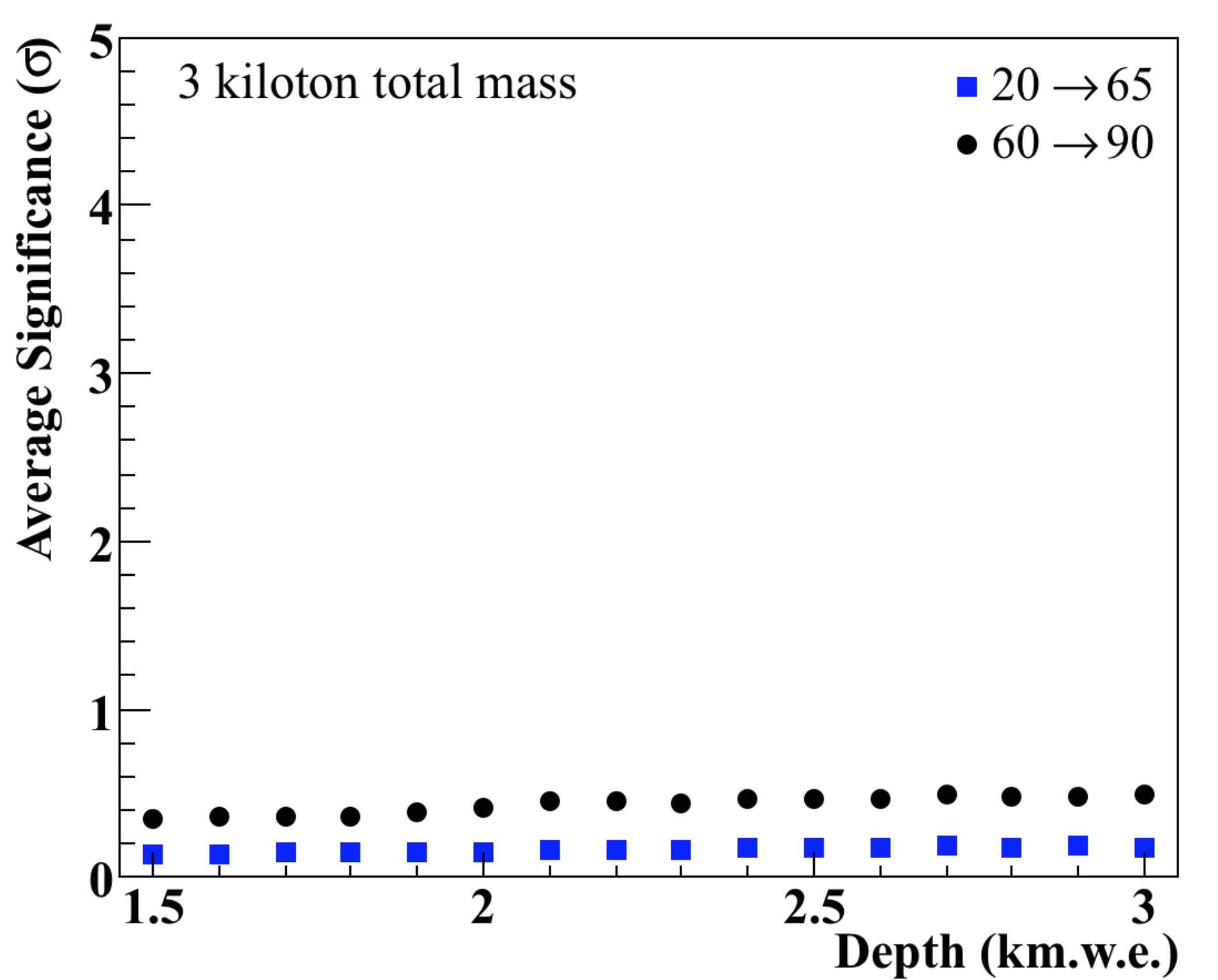} & 
\includegraphics[height=0.255\linewidth]{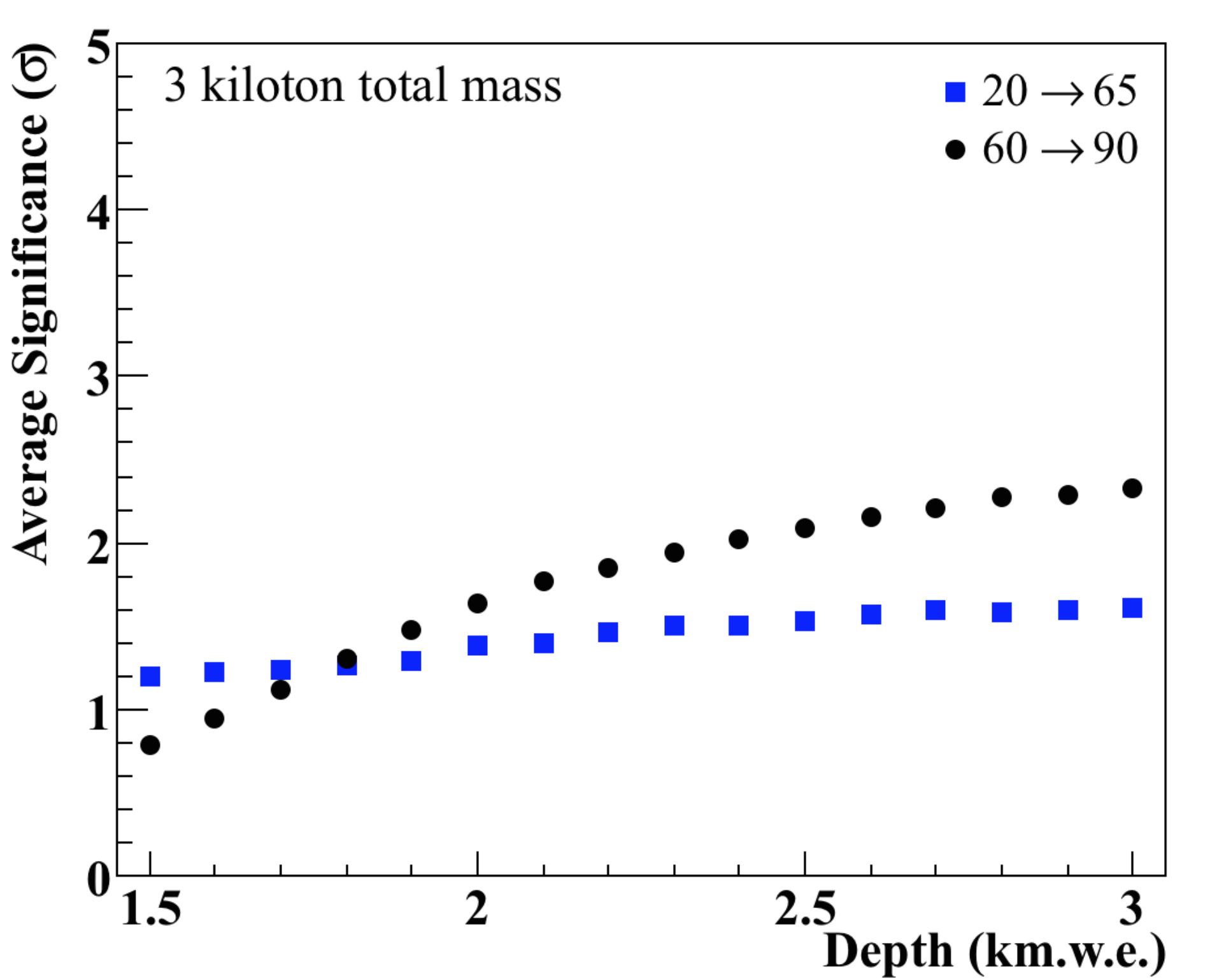} &
\includegraphics[height=0.255\linewidth]{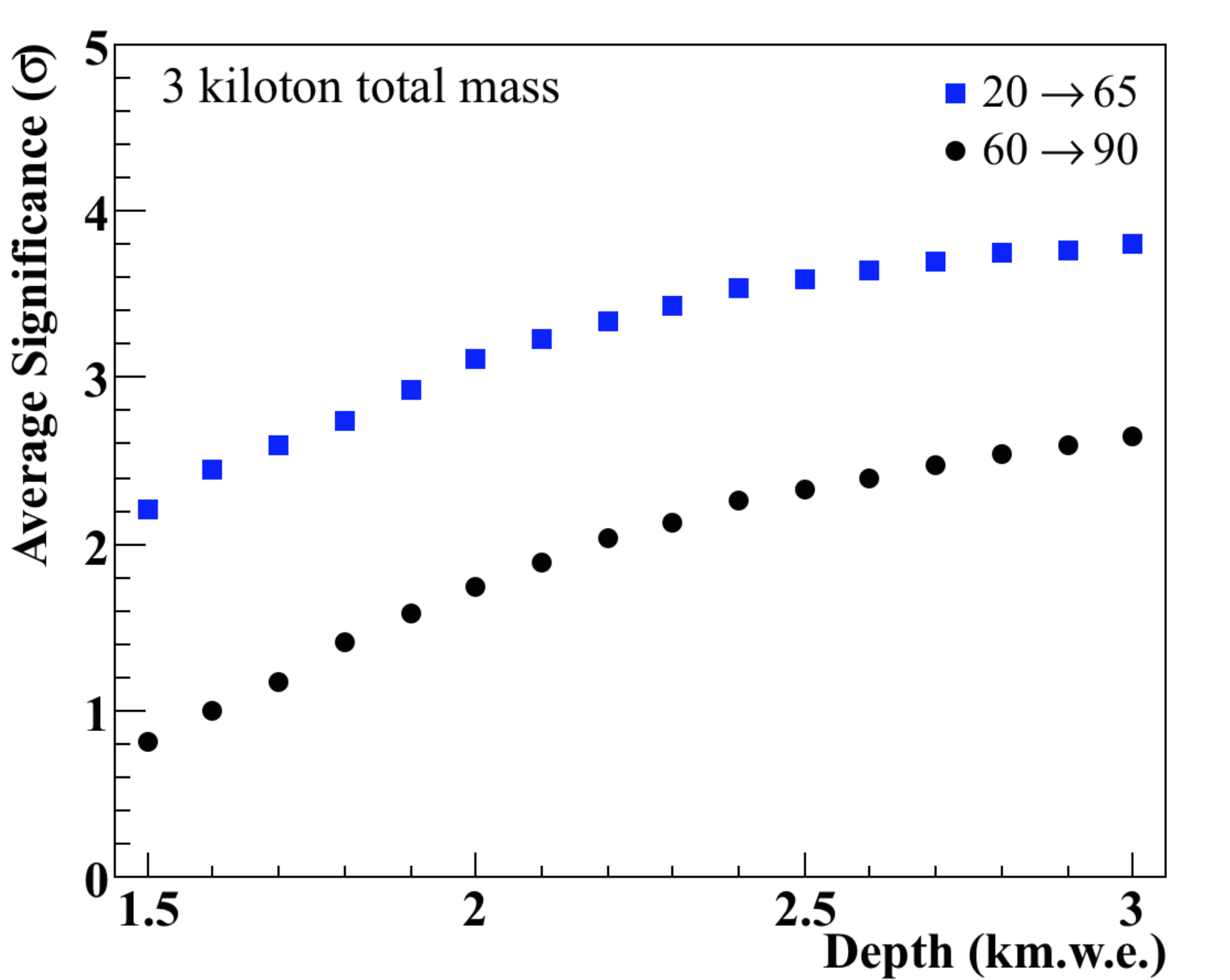} 
\\
\footnotesize (a) & \footnotesize (b) & \footnotesize (c)
\\
\\
\includegraphics[height=0.255\linewidth]{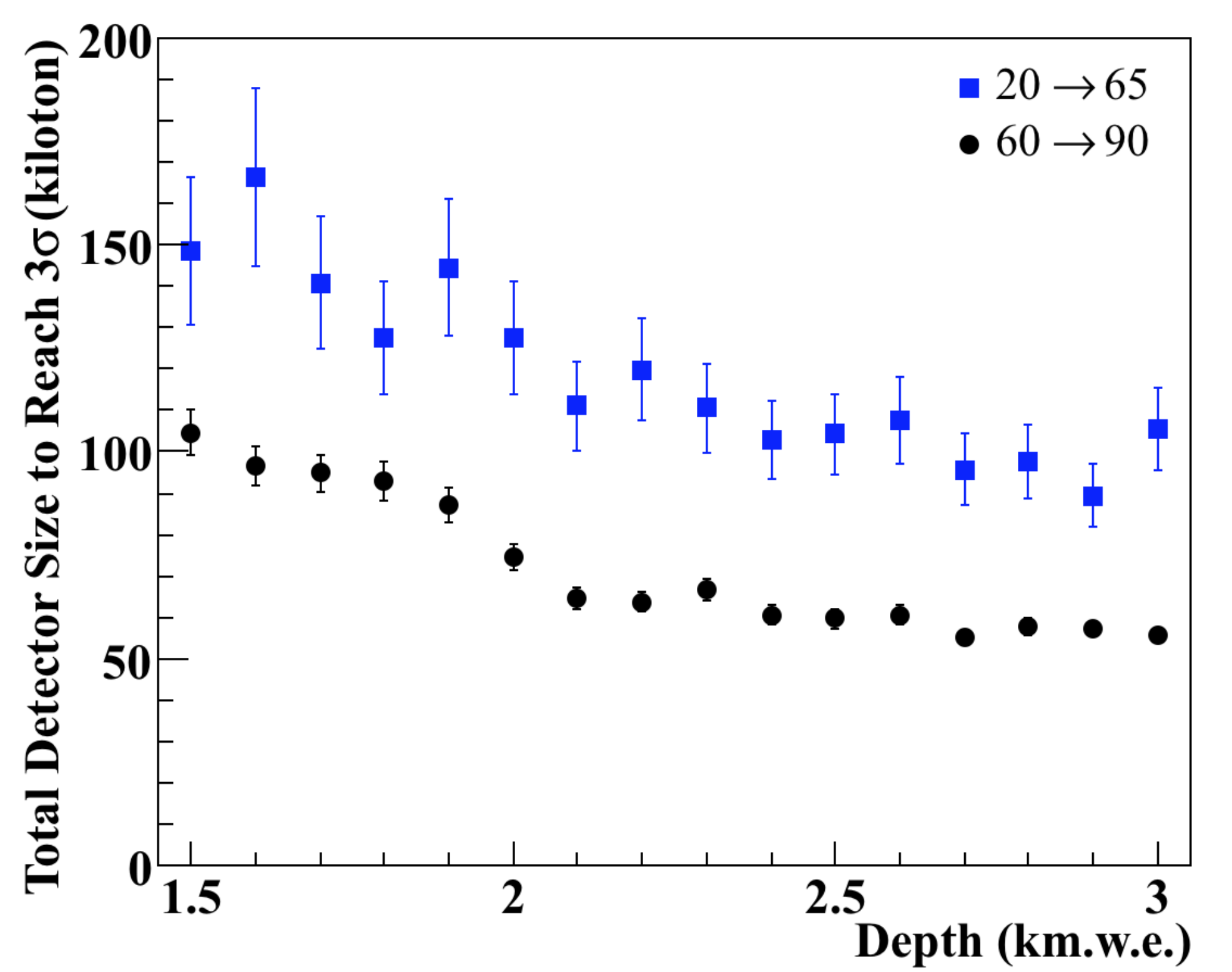} & 
\includegraphics[height=0.255\linewidth]{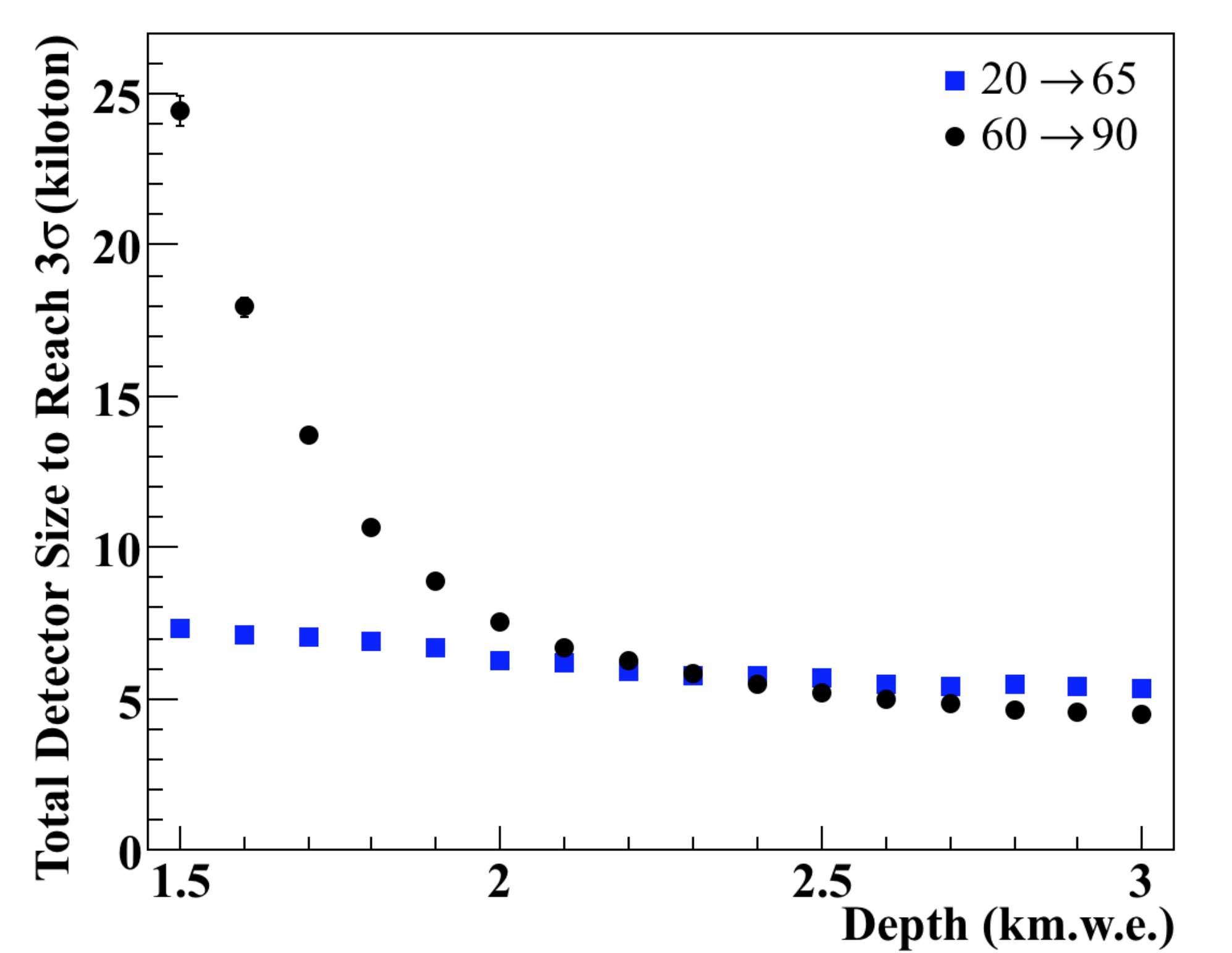} & 
\includegraphics[height=0.255\linewidth]{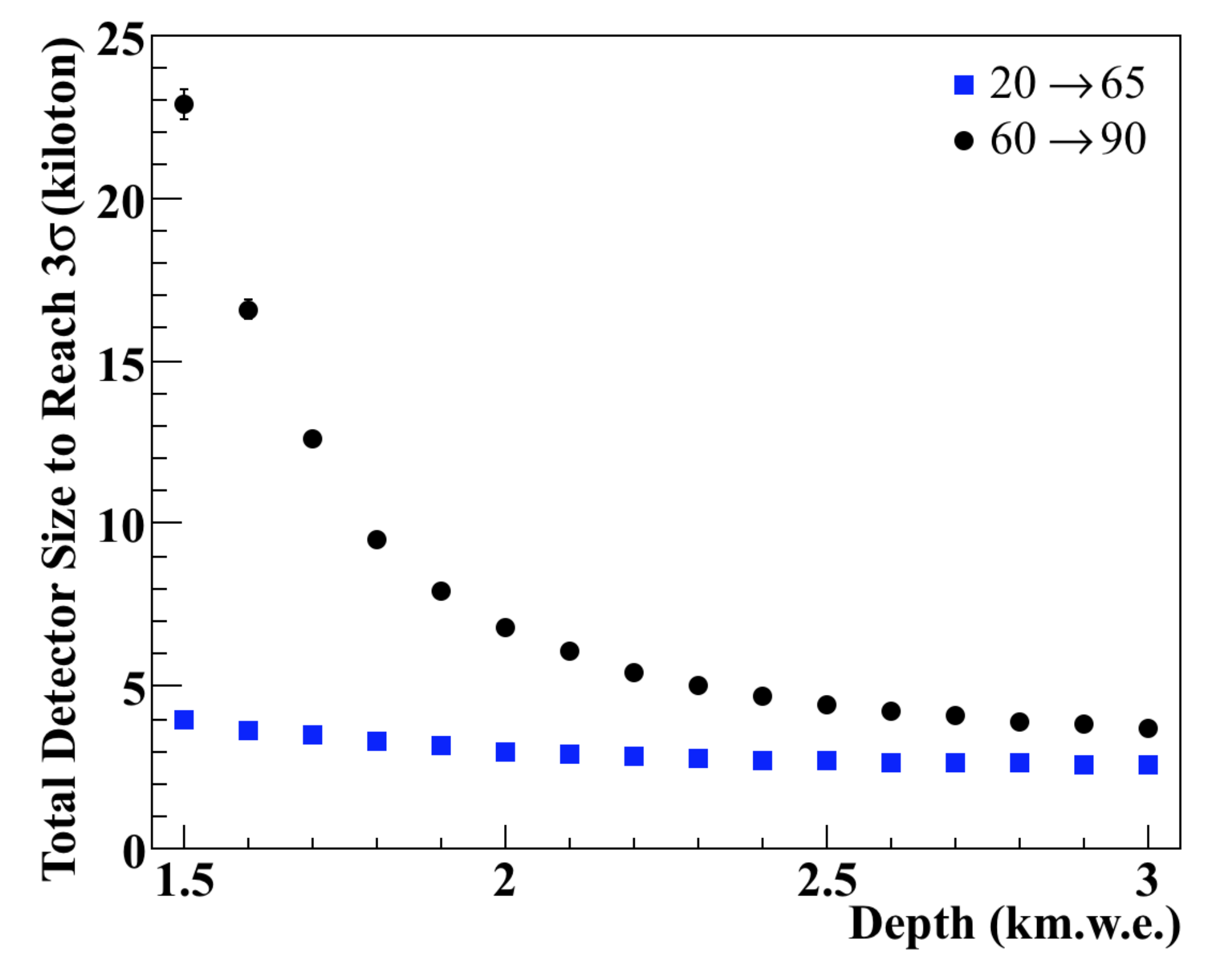}
\\
\footnotesize (d) & \footnotesize (e) & \footnotesize (f)
\end{tabular}
\caption{Average statistical significance in a 3-kiloton detector (total mass) plotted as a function of depth using two different energy ranges considered here (25 to 65 and 60 to 90 hit PMTs), with radon levels of 1 $\times$ SNO (a), 10\textsuperscript{-2} $\times$ SNO (b), and 10\textsuperscript{-4} $\times$ SNO (c). Error bars are included and represent the uncertainty in the mean (however they are too small to be seen in most cases). The uncertainty in a single experiment is $\pm$1$\sigma$. Total detector size required for 3$\sigma$ significance plotted as a function of depth for radon levels of 1 $\times$ SNO (d), 10\textsuperscript{-2} $\times$ SNO (e), and 10\textsuperscript{-4} $\times$ SNO (f). Again, error bars represent the uncertainty in the mean and do not represent the uncertainty in a single experiment. Therefore the results should only be used as a guide.}
\end{figure*}

If in fact radon levels cannot be reduced relative to SNO, the detector size needs to be increased significantly ($>$ 50 kilotons) in order for directionality to be possible. If significant radon reduction is possible, detector sizes anywhere from 3 kilotons (WATCHMAN-size) to 10 kilotons may be directionally sensitive, depending on the specific depth and radon levels.

\section{\label{sec:conclusions}Conclusions}

Our study shows that under certain conditions, the reconstruction of the direction of a reactor may be achievable via the antineutrino-electron scattering channel. The main factors affecting sensitivity are radon contamination and overburden.  With similar water-borne radon levels to SNO and a 3000-m.w.e.\,\,depth, a fairly large detector approximately the size of Super-K (55-kiloton total, 37-kiloton fiducial) is required for 3$\sigma$ sensitivity to the assumed 3.758-GWt reactor at 13-km standoff. With a factor of 100 radon reduction and at least 2000-m.w.e.\,\,depth, a more tractable 6.3-kiloton (885-ton fiducial) detector is sufficient for the same reactor power and significance level (see \hyperlink{appendixB}{Appendix B} for an example of a directional plot with these conditions). If a significant reduction in radon is possible (10\textsuperscript{-4} $\times$ SNO), a 4-kiloton (172-ton fiducial) detector at a shallower 1500-m.w.e.\,\,depth (similar to WATCHMAN) would be directionally sensitive for the same reactor power and significance level. Assumptions in these results include similar steel cleanliness levels to the KamLAND detector, a continuously operated reactor at full power with no shutdown periods, and constant fission fractions typical of a mid-cycle PWR. Furthermore, the situation investigated here is the directional sensitivity of an incoming antineutrino flux with respect to an assumed reactor location. If the true location is unknown, a statistical penalty would need to be applied for testing in multiple directions.

More generally, flux scaling allows us to approximate directional sensitivity at greater distances and for smaller reactor power levels. Assuming the case of 10\textsuperscript{-2} $\times$ SNO radon contamination and 2500-m.w.e.\,\,overburden, directional reconstruction of a 3.758-GWt reactor at 3$\sigma$ significance would be possible at a 70-km standoff with a 1-megaton (757-kiloton fiducial) detector. Equivalently, a megaton detector would be sensitive to a 125-MWt reactor at 13 km. If the radon contamination is reduced by 10\textsuperscript{4}, the 3.758-GWt reactor standoff increases to 105 km while the smallest detectable reactor at 13 km decreases to 55 MWt. Megaton-sized water-based detectors represent the outer limit of what is possible in field-able neutrino detectors \cite{Abe3}.

While these conditions may be difficult to achieve in practice, we have demonstrated that Gd-doped WCDs have the potential to utilize elastic electron scattering for nuclear reactor antineutrino directionality. Compared with the WATCHMAN detector, the main factors for improvement in directional pointing are greater depth and 10\textsuperscript{-2} or less radon contamination in the inner detector volume compared with the SNO detector. We hope that this research may serve as a catalyst to pursue an R\&D effort into water-borne radon removal techniques for future large scale Gd-doped WCDs used for remote monitoring of nuclear reactors.

\hypertarget{table3}{}
\begin{table*}[!htbp]
\setlength{\tabcolsep}{7pt}
\centering
\caption{Signal and background estimates for a WATCHMAN-like detector at a depth of 2350 m.w.e\,\,for 5 years assuming two different energy analysis cuts. Average significances were calculated assuming the radon levels relative to those of SNO. The radionuclide background is denoted by ``RN" and the backgrounds due to steel, rock, misidentified IBD, and solar neutrinos are combined together and denoted by ``Other". Since the ideal FV can change with increasing energy, we include the range of FVs used within each energy slice.} 
\begin{tabular}{ c  c  c  c | c  c  c  c  c  c  c }
\multicolumn{11}{c }  {5 Year Acquisition}  \\ 
\hline 
\multirow{2}{*} {\begin{tabular}{c} {PMT} \\ {Triggers} \end{tabular} } &
\multirow{2}{*} {\begin{tabular}{c} {FV} \\ {(m\textsuperscript{3})} \end{tabular}} &
\multirow{2}{*} {ES} &
\multirow{2}{*} {\begin{tabular}{c} {Exp.} \\ {Slope} \end{tabular} } &
\multirow{2}{*} { } &
\multirow{2}{*} {RN} &
\multirow{2}{*} {PMTs} &
\multirow{2}{*} {Other} &
\multicolumn{3}{c } {\multirow{1}{*} {Radon ($\times$ SNO)}}  \\ \cline{9-11} 
&&&&&&&& \multirow{1}{*} {1} & \multirow{1}{*} {10\textsuperscript{-2}} & \multirow{1}{*} {10\textsuperscript{-4}} \\
\hline
25 $\rightarrow$ 65 & 187 & 97 & 4.6 & Bkgd.\,\,Components & 123 &  1463 & 511 & 1148920 & 11489 & 115 \\
&&&&&&&&&& \\ [-2.5ex] \cline{5-11}
& & & & Total Background & & & & 1151017 & 13586 & 2212 \\
& & & & Significance & & & & 0.1$\sigma$ & 1.5$\sigma$  & 3.5$\sigma$ \\ [1ex]

60 $\rightarrow$ 90 & 500 - 1000 & 51 & 6.7 & Bkgd.\,\,Components & 722 & 270 & 1278 & 61485 & 615 & 6 \\
&&&&&&&&&& \\ [-2.5ex] \cline{5-11}
& & & & Total Background & & & & 63755 & 2885 & 2276 \\
& & & & Significance & & & & 0.5$\sigma$ & 2.0$\sigma$ & 2.2$\sigma$ \\ [1ex]
\hline 
\end{tabular}
\end{table*}

\section*{\label{sec:acknowledgements}Acknowledgements}
 
The authors would like to thank Marc Bergevin of Lawrence Livermore National Laboratory for his help with the development and usage of RMSim for the purposes of this work and Michael Smy of the University of California, Irvine for his assistance with BONSAI. This work was supported by the U.S. Department of Energy National Nuclear Security Administration [Award No.\,\,DE-NA0000979] through the Nuclear Science and Security Consortium, and Lawrence Livermore National Laboratory [Contract No.\,\,DE-AC52-07NA27344, LDRD tracking number 15-ERD-021, release number LLNL-JRNL-679610]. The views and opinions of authors expressed herein do not necessarily state or reflect those of the United States Government or any agency thereof.

\hypertarget{appendixA}{}
\section*{\label{app:appendixA}{Appendix A. Example: Signal and Background Estimates at Kamioka Depth}}

Table \hyperlink{table3}{III} shows the predicted number of reactor-induced elastic scattering events 13 km from a 3.758-GWt power reactor, and the expected number of background events after 5 years in a WATCHMAN-like detector at the same depth as the KamLAND detector (2350 m.w.e.). ``WATCHMAN-like" refers to a $0.1\%$ Gd-doped water Cherenkov detector with a total detector volume of just over 3 kilotons, a 2-kiloton inner detector volume and a nominal 1-kiloton fiducial volume. The actual fiducial volume depends on the energy cuts used. The two energy regimes used here extend from 25 to 65 triggered PMTs and from 60 to 90 triggered PMTs. The table also includes the fixed exponential slope used in Eq.\,(\hyperlink{eq9}{9}) and the average statistical significance for the three different assumed radon contaminations relative to the SNO detector.

\newpage
\hypertarget{appendixB}{}
\section*{\label{app:appendixB}{Appendix B. Example: Directional Signal and Background Plot}}

Figure \hyperlink{fig9}{9} shows an example 5-year directional reconstruction of all signal and background events that trigger between 25 and 65 PMTs in a 6.3-kiloton (885-ton fiducial) detector at a depth of 2000 m.w.e.\,\,with radon contamination reduced by a factor of 100 relative to SNO. These conditions represent a more tractable experimental design option to achieve a 3$\sigma$ directional signal with respect to an assumed known direction.

\begin{figure}[!htb]
\hypertarget{fig9}{}
\centering
\includegraphics[width=225pt]{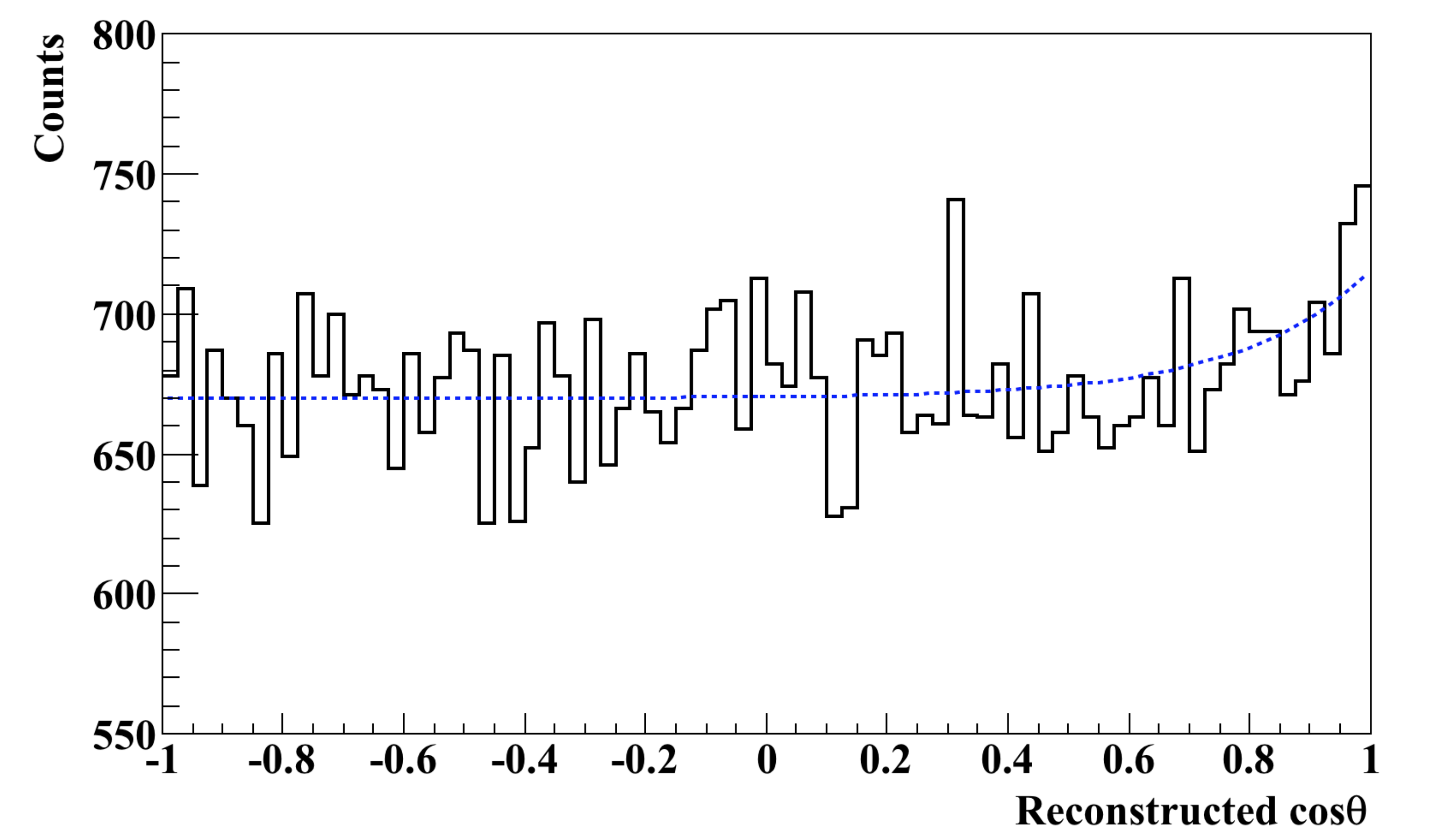}
\caption{(Color online) Cosine of the reconstructed angle for all low-energy (25 to 65 triggered PMTs) events in 5 years for a 6.3-kiloton (885-ton fiducial) detector at a depth of 2000 m.w.e.\,\,with a radon level of 10$^{-2}$ $\times$ SNO. The exponential fit shown in blue has a fixed slope of 4.6 as in Table \protect\hyperlink{table3}{III}. The directional significance of the antineutrino source at $\cos\theta = 1$ from this particular data is $3.1\sigma$. }
\end{figure}

\bibliography{Directionality}

\end{document}